# In-situ visualization of local distortions in the high-$T_c$ molecule-intercalated $Li_x(C_5H_5N)_yFe_{2-z}Se_2$ superconductor

Izar Capel Berdiell,[1] Edyta Pesko,[2] Elijah Lator,[1] Alexandros Deltsidis,[1,3] Anna Krztoń-Maziopa,[2] A. M. Milinda Abeykoon,[4] Emil S. Bozin,[5] Alexandros Lappas,[1,*]

[1] *Institute of Electronic Structure and Laser, Foundation for Research and Technology–Hellas, Vassilika Vouton, 711 10 Heraklion, Greece.*

[2] *Faculty of Chemistry, Warsaw University of Technology, 00-664 Warsaw, Poland.*

[3] *Department of Materials Science and Technology, University of Crete, Voutes, 71003 Heraklion, Greece.*

[4] *Photon Sciences Division, National Synchrotron Light Source II, Brookhaven National Laboratory, Upton, NY 11973, USA.*

[5] *Condensed Matter Physics and Materials Science Division, Brookhaven National Laboratory, Upton, New York 11973, USA.*

[*] e-mail: lappas@iesl.forth.gr

## ABSTRACT

A time-resolved synchrotron X-ray total scattering study sheds light on the evolution of the different structural length scales involved during the intercalation of the layered iron-selenide host by organic molecular donors, aiming at the formation of the expanded lattice $Li_x(C_5H_5N)_yFe_{2-z}Se_2$ hybrid superconductor. The intercalates are found to crystallize in the tetragonal $ThCr_2Si_2$-type structure at the average level, however, with an enhanced interlayer iron-selenide spacing (d= 16.2 Å) that accommodates the heterocyclic molecular spacers. Quantitative atomic pair distribution function (PDF) analysis at variable times, suggests distorted $FeSe_4$ tetrahedral local environments that appear swollen with respect to those in the parent *β*-FeSe. Simultaneously acquired, in-situ synchrotron X-ray powder diffraction data disclose that secondary phases (*α*-Fe and $Li_2Se$), grow significantly when higher lithium concentration is used in the solvothermal reaction or when the solution is aged. These observations are in line with the strongly reducing character of the intercalation medium's solvated electrons that mediate the defect chemistry of the expanded lattice superconductor. In the latter, intralayer correlated local distortions indicate electron donating aspects that reflect in somewhat enlarged Fe-Se bonds. They also reveal a degree of relief of chemical pressure associated with a large distance between Fe and Se sheets ('taller' anion height) and a stretched Fe-Fe square planar topology. The elongation of the latter, derived from the in-situ PDF study, speaks for a plausible increase in the Fe-site vacancy concentration. The evolution of the local structural parameters suggests an optimum reaction window where kinetically stabilized phases resemble the distortions of the edge-sharing Fe-Se tetrahedra, required for high-$T_c$ in expanded lattice iron-chalcogenides.

**Keywords:** iron-based superconductors, intercalation, local structure, iron-vacancy, defect chemistry

## INTRODUCTION

Iron-based superconductors involve two broad families of correlated electron systems identified by the type of anion, namely, pnictogen (Pn= As, Sb) or chalcogen (Ch= Se, Te), that facilitate two-dimensional layers where iron has a four-fold coordination. While the Fe-Ch layers maintain a charge neutrality,[1] the different nature of the anion in the pnictides allows for negative charge in the Fe-Pn layers. In this respect, superconductivity in the latter requires the insertion of cations (e.g. LiFeAs)[2] or PbO-type oxide spacers (*Ln*OFeAs; *Ln*= lanthanides)[3,4] for charge balancing purposes. In the former, though, stoichiometry is necessary for the emergence of superconducting behavior, and as such the defect chemistry of the Fe-Ch layers plays a pivotal role. Electronic properties in iron-superconductors are then mediated by modifications[5] involving the in-plane $Fe(Pn/Ch)_4$ edge-sharing tetrahedral building blocks.

Among the iron chalcogenides, the binary $\beta$-$FeSe_{1-x}$ layered material has attracted great attention as it shows a 8 K superconducting transition ($T_c$),[1] a property that has motivated intense interest since it is accompanied by a dramatic pressure-induced enhancement[6,7] up to 37 K and a boost from 32 K to 46 K, when the $\beta$-FeSe layers become intercalated by alkali or alkaline- or rare- earth metals (cf. $AFe_{2-y}Se_2$ phases; A= Li, Na, Tl/K or Sr, Ba or Yb, Eu, etc.).[8] The $T_c$ improvement has been correlated with the interplay between the concomitant modification of the interlayer separation and the charge doping induced changes of the Fe-selenide electronic structure imposed by the electron-donating capabilities of the intercalants (guest species). The unprecedented $T_c$ enhancement to about 56 K in the single-layer FeSe/ $SrTiO_3$ interfacial superconductor,[9] establishes the complex interplay of the pairing mechanism and the reduced dimensionality in such a family of Fe-chalcogenide two-dimensional (2D) materials.[10] In this context, the origin of superconducting mechanism and the required composition that yields maximal $T_c$ are currently under debate.[11]

However, in the race to discover single-phase materials that operate at elevated $T_c$, high-temperature solid-state synthesis of the $AFe_{2-y}Se_2$ series, consistently resulted in detrimental microscale phase separation.[12] While the coexisting phases appear to be charge balanced, with formally $Fe^{2+}$ valence state present, a variable degree of crystal defects at the iron and alkali metal sites appears unavoidable. For example, in the K-Fe-Se phase space, a minority superconducting phase ($K_xFe_{2-y}Se_2$) is inter-grown coherently with a majority antiferromagnetic semiconductor phase ($K_2Fe_4Se_5$) that involves ordered vacancies.[13] In contrast to the parent *β*-$FeSe_{1-x}$, where a narrow composition window (δ= 0.01-0.03, x< 0.03) stabilizes bulk superconductivity,[14,15] the controlled defect chemistry of intercalated Fe-Ch phases is a critical parameter for the adjustment of their electronic properties.[16]

In view of such a demand, solvothermal routes, where alkali metals were dissolved in ammonia or other electron-rich donor molecules (cf. organic amines), in analogy to well-exploited transition metal dichalcogenides,[17,18] are explored as promising low-temperature alternatives. In this direction, molecules co-intercalated with alkali metals, in the van der Waals interlayer gap of the pre-made host material *β*-FeSe, offer a viable synthesis route, where the amount of Fe-vacancies is seemingly a control parameter. The derived phases, $A_x(molecule)_yFe_{2-z}Se_2$ (A= alkalis; molecules= $NH_3$,[19] heterocyclic pyridine,[20] alkyl amine,[21–23] aliphatic diamines,[24] etc.), grown ammonothermally or even hydrothermally, contain alternating layers of *β*-FeSe, with slabs of $Li_x(NH_2)_y(NH_3)_{1-y}$,[19] $Li_x(C_5H_5N)_y$,[20] or even $(Li_{1-x}Fe_x)OH$[25] intercalated within the interlayer space. As the inclusion of guests [alkali metal - molecule] increases the interlayer separation and leads to an important fourfold rise of the $T_c$ (40-45 K), their presence motivates important questions related to the role of intercalated species in promoting the charge doping of the electronically active FeSe layers, and enquire how this impacts the evolution and magnitude of $T_c$.

In such molecule intercalated phases,[26] low-temperatures and slowed-down kinetically driven processes in liquid media, may affect not only the crystallinity of the as-made products[24] but can also arrest parasitic phases, leading to intermittent superconducting behavior. To characterize this, time-resolved in-situ synchrotron X-ray and neutron diffraction, have been successfully implemented for phase identification and structure determination, on the course of the ammonothermal synthesis of alkali[27] and alkaline-earth[28] intercalates of *β*-FeSe. In an effort to develop new opportunities for optimal growth conditions that can address inadequacies in the "A-amine-FeSe" phase field, the in-situ studies uncovered vital information on the role of $T_c$-scaling parameters, such as the interlayer spacing, the modification of the local structure based on the $FeSe_4$ tetrahedra and their association with the degree of the carrier doping.

However, among organic molecular hybrid superconductors with large heterocyclic amine spacers and alkali metals intercalated in *β*-FeSe, those in the $Li_x(C_5H_5N)_yFe_{2-z}Se_2$ ($C_5H_5N$= PyH5= Pyridine) series are much less explored. Understanding such complex systems faces a twofold challenge. On the technical side, limitations to synthesis that preclude a pure product, potential presence of intermediate phases and their detailed nature need to be established. It is important to determine if such components can be manipulated by optimizing the synthesis parameters, such as reaction temperature, time, molarity etc.. On the science side, identification of factors that are responsible for the $T_c$ enhancement, such as the layer isolation, or the electronic doping, or their combination, is essential for further optimization of superconductivity in these materials. In this endeavor, we designed a portable reactor setup and employed it in high-throughput, high-energy, time-resolved synchrotron X-ray total scattering to track simultaneously the changes in the local and average structures of the *β*-FeSe host during its intercalation by [Li-PyH5]. Total scattering combined with pair distribution function (PDF) analysis is the method of choice, because uncovering the local structure gives different perspectives, such as whether the intercalation proceeds locally first and when long range order is established, as these spatial degrees of freedom could potentially evolve at different time scales. In addition, when it comes to doping and its impact on the crystal structure, this is expected to be most reliably seen in the changes of local environments (cf. $FeSe_4$ units). The work provides a systematic overview of the intercalation process, revealing reaction variables and formation of competing phases, thus addressing the efficacy of synthesis and helping its further optimization to stabilize superconductivity in expanded-lattice, electron-doped iron-chalcogenide materials.

## EXPERIMENTAL METHODS

Reference polycrystalline parent tetragonal *β*-FeSe and hexagonal *δ*-FeSe polymorphs were prepared following standard solid-state chemistry protocols.[15] Synthesis of powders from the $Li_x(C_5H_5N)_yFe_{2-z}Se_2$ intercalates utilized low-temperature solvothermal methods.[20] All sample manipulations, were undertaken inside an Ar-circulating MBRAUN (UNILab) glove box. Phase purity of the prepared materials, enclosed in an air-tight samples holder, was tested with an X-ray powder diffractometer (BRUKER D8 ADVANCE), with Cu-$K_\alpha$ radiation ($\lambda$ = 1.5418 Å). Where necessary, temperature dependent AC magnetic susceptibility measurements ($H_{ac}$= 1 Oe and f= 999 kHz), were obtained by utilizing an Oxford Instruments MagLab EXA 2000 vibrating sample magnetometer (ESI §S1).

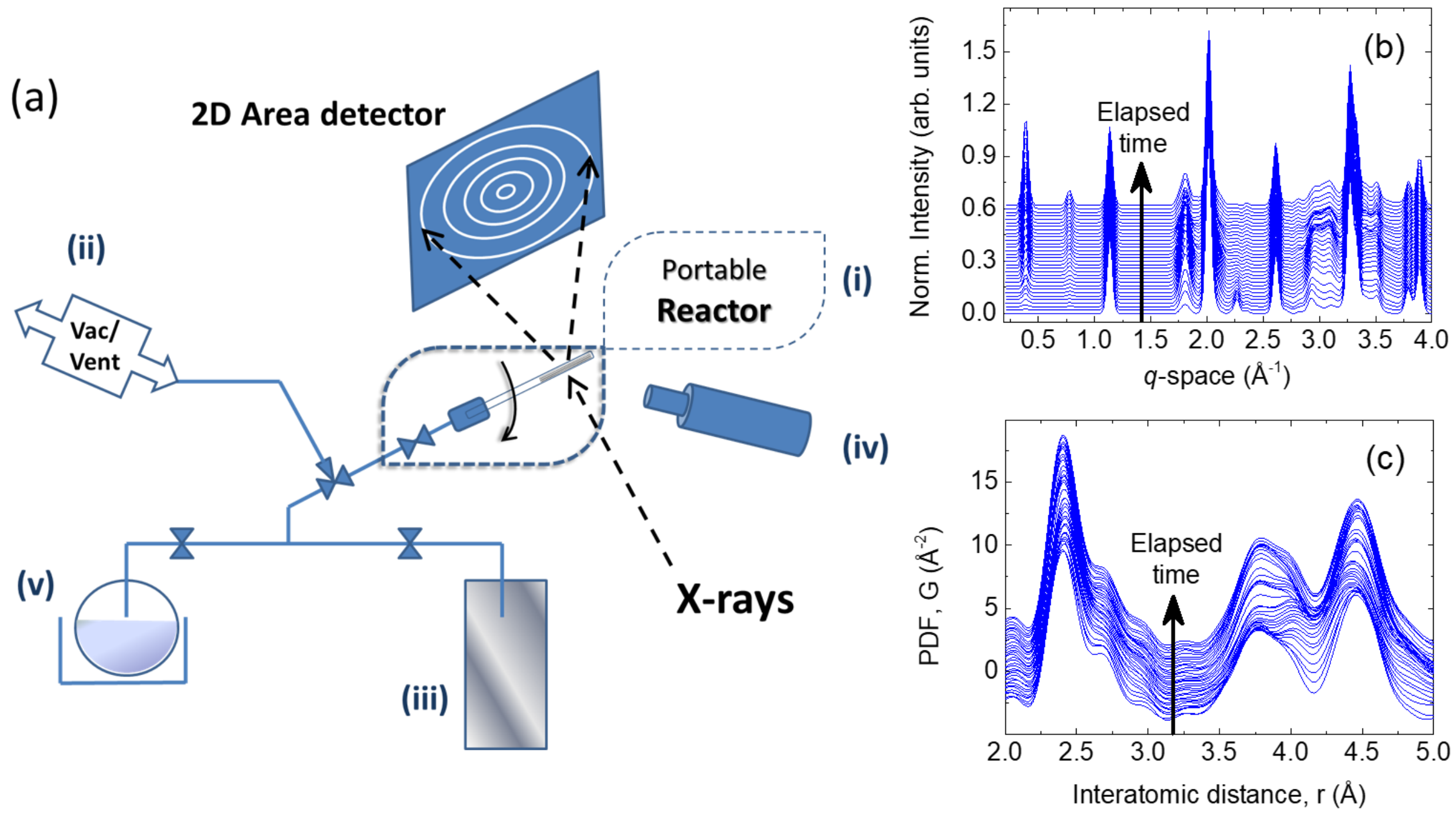

**Figure 1.** (a) Schematic, depicting the setup for an in-situ synchrotron X-ray total scattering study at the 28-ID-1 beamline. A portable spinning reactor (i) was developed for studying the real-time reactivity of *β*-FeSe layers in liquid-media (Fig. S1, §S1 ESI). The isopiestic technique enabled the transfer of molecular spacers [pyridine] to the [Li:FeSe] reagents, with the support of (ii) vacuum pump, (iii) He-gas supply, (iv) liquid $N_2$ cryostream, for sample environment temperature control, (v) pyridine container, equipped with heating mantle. Representative time-resolved synchrotron XRD (b) and PDF (c) data collected simultaneously. With the sample-to-detector at 'near' side, they depict structure changes at the average and local length scales in the Li-PyH5-FeSe phase field (M= 0.7, experiment (**C**)).

Synchrotron X-ray total scattering measurements were carried out with a monochromatic 74.375 keV (λ = 0.1667 Å) X-ray beam at the 28-ID-1 beamline at the National Synchrotron Light Source-II of Brookhaven National Laboratory. A fast-responding two-dimensional (2D) PerkinElmer area detector was used, in two different sample-to-detector positions, namely, at 326 mm (near) and 1124 mm (far). This enables either a broad q-range appropriate, for pair distribution function (PDF) data acquisition or higher-angular resolution, suitable for X-ray powder diffraction (XRD) measurements. The PDF analysis was carried out with the PDFgui[29] package. The powder XRD patterns were examined by LeBail-type full profile analysis using the FullProf[30] suite. For the in-situ X-ray total scattering study of the $Li_x(C_5H_5N)_yFe_{2-z}Se_2$ phase formation, a portable reactor (Fig. S1) was developed and integrated into the 28-ID-1 experimental setup, allowing us to probe the real-time reactivity of $\beta$-FeSe layers in liquid-media (Fig. 1). As the intercalated materials' growth is regulated by varying the molarity (M) of the [Li:PyH5] solution, within the set time-frame of the synchrotron facility study, three in-situ experiments assuming, freshly made 0.7 M (**A**), 1.4 M (**B**), as well as 72 h aged 0.7 M (**C**) solutions, were deployed to probe the synthesis of $Li_x(C_5H_5N)_yFe_{2-z}Se_2$ at 80°C. The sample temperature was controlled with the use of $N_2$ cryostream. The time-resolved processes were then studied not only with medium q-space resolution to study the longer length scales (Fig. 1b; e.g. interlayer separation during specimen intercalation), but importantly with sufficient r-space resolution ($q_{max}$~24 Å$^{-1}$) to explore local structure modifications (Fig. 1c; e.g. due to specimen doping).

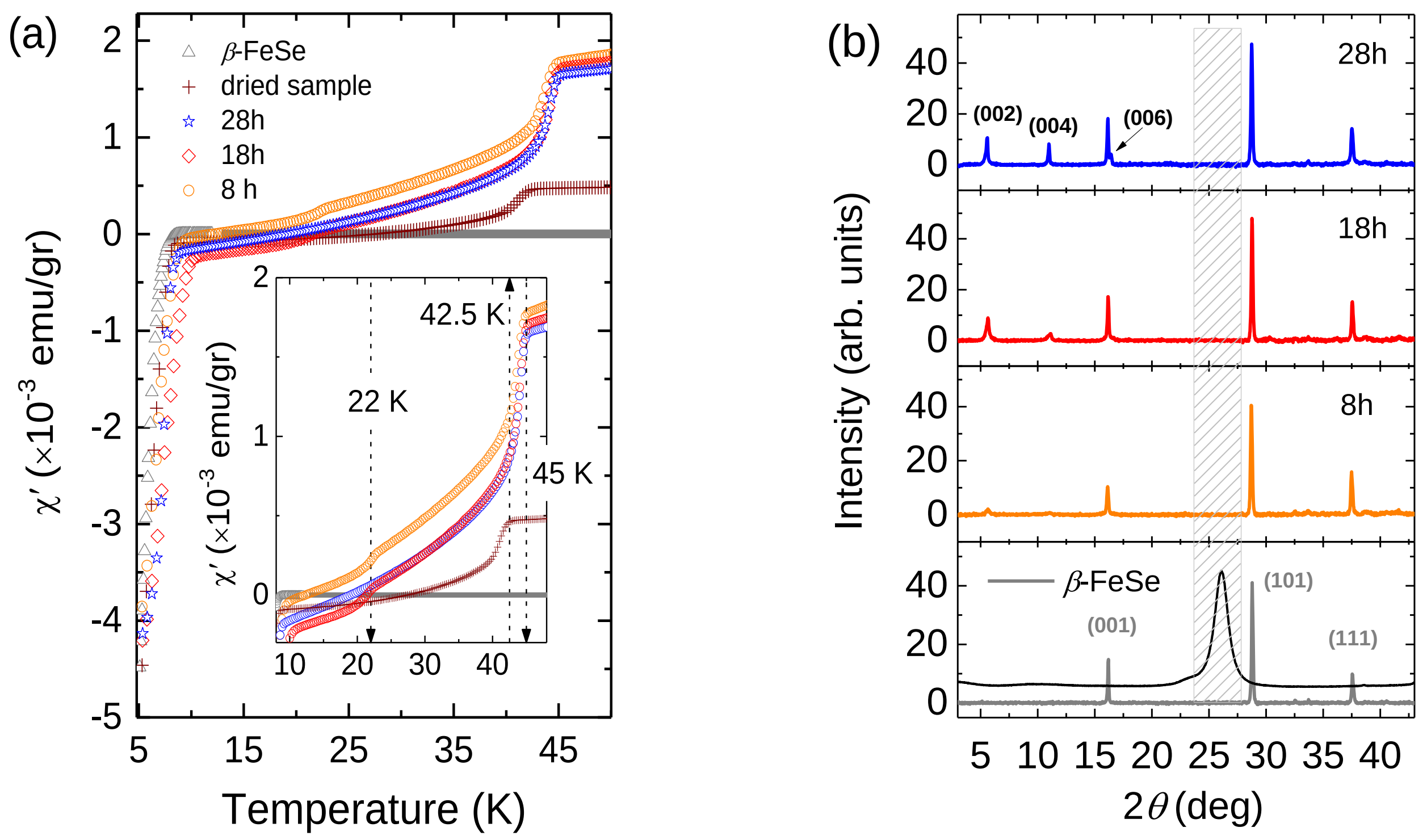

**Figure 2.** (a) AC susceptibility of $Li_x(C_5H_5N)_yFe_{2-z}Se_2$ samples formed in liquid media (wet samples), flamed-sealed inside quartz ampoules; inset: magnified temperature region near $T_c$, and final dried sample (+); $H_{ac}$= 1 Oe, f= 999 Hz. (b) powder XRDs (CuKa) of the corresponding dried specimen, at different times of the intercalation reaction. The collected set of data correspond to $\beta$-FeSe starting material and the 8 h (grey), 18 h (orange) and 28h (orange) $Li_x(C_5H_5N)_yFe_{2-z}Se_2$ (nominal x~1.0) intercalates. Shaded XRD area, 2θ= 24-28°, is an excluded region due to contribution from a home-made, air-tight sample holder; black line profile, over plotted with the $\beta$-FeSe pattern – signal is offset for clarity.

## RESULTS & DISCUSSION

**Ex-situ observations.** To follow up on earlier observations where low-T solvothermal reactions (40-80°C, over 24 hr) produce materials with low superconducting volume fractions,[20] we also reacted *β*-FeSe with a [Li:PyH5] solution (M= 0.6) reproducing the aforementioned conditions. The properties of the ex-situ formed products were probed by collecting magnetic susceptibility data on wet samples, at various time intervals, namely, at 8, 18 and 28 h, as well as on a dried specimen (Fig. 2a). It is apparent that the 8 K component, belonging to the starting material, is more pronounced than the transition from the expanded-lattice phase ($T_c$~ 45 K) even at long reaction times (28 h). The presence of the starting material is also confirmed via XRD by the intense (101) Bragg peak, typical of *β*-FeSe (Fig. 2b), which coexists throughout the reaction times with the growing (00*l*) reflections of the intercalated phase. Small transitions in the susceptibility at ~22 K, are evident at the 8 and 18 h time intervals, but not for the 28 h experiment. Such an "intermediate" phase might be related to the presence of a meta-stable, lower electron doped material that evolves into the main intercalated phase at longer reaction times. The dried sample measurement implies that the superconducting volume fraction of the intercalated phase decays to half upon drying. With its $T_c$ shifting to ~42 K, it is also worth noting the magnitude of the paramagnetic moment which drops to roughly one third, but remains present, inferring parasitic phases that may interfere with superconductivity (*vide-infra*).

In view of the complex behavior encountered so far, implying a sub-optimal synthesis procedure, with issues on reproducibility, we employed the brilliance of synchrotron X-ray powder diffraction to study the room-temperature average structure of a dried $Li_x(C_5H_5N)_yFe_{2-z}Se_2$ (nominal x~2.0) material, prepared in advance. Indexing of the XRD pattern (Fig. 3a) and its LeBail full profile analysis (Fig. S3a), indicate that in typical, ex-situ grown samples the expanded lattice 42 K superconducting phase, coexists with other phases, including, *α*-Fe, $Li_2Se$, and *β*-FeSe (tetragonal – P4/nmm). Here, the basic structure of the intercalated phase has an interlayer FeSe distance of d= 16.2 Å, as determined by the low-angle (00*l*) Bragg reflection (q= 0.39 $Å^{-1}$, 2θ= 0.55°, λ = 0.1667 Å). Adjacent layers though, while separated by electron-donating spacers, can be either co-aligned out-of-plane (cf. PbFCl, 111-type [31,32]; Fig. 3c) or alternating in orientation ($ThCr_2Si_2$, 122-type [33,34]; Fig. 3d) along the c-axis. In view of these topologies, we simulated the powder patterns of 111 and 122 types of FeSe frameworks and compared them with the observed experimental data (Fig. 3b). This qualitative comparison of simulated and experimental diffraction profiles, suggests that the $Li_x(C_5H_5N)_yFe_{2-z}Se_2$ may be indexed on the basis of the I4/mmm symmetry (a=b≅ 3.8 Å , c≅ 32.47 Å) met in the $ThCr_2Si_2$ structure type (ESI §S3).

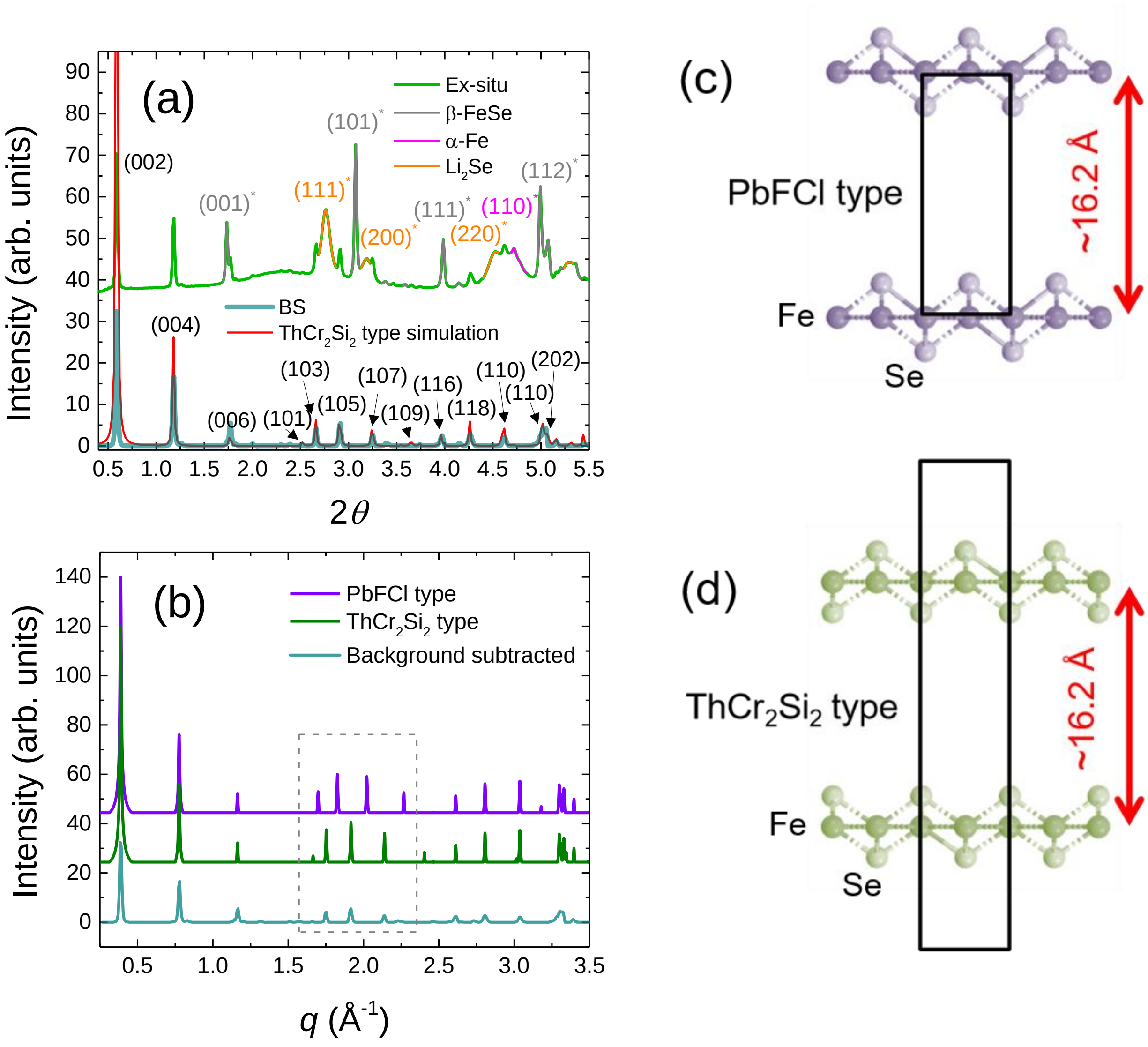

**Figure 3.** (a) Synchrotron X-ray powder diffraction pattern (λ= 0.1667 Å; sample-to-detector position at 'far' side) of an ex-situ grown sample, where together with the expanded-lattice $Li_x(C_5H_5N)_yFe_{2-z}Se_2$ (nominal x~2.0) reflections, a number of secondary phases are identified, namely, $\alpha$-Fe, $Li_2Se$, $\beta$-FeSe (tetragonal – P4/nmm), (b) simulated powder X-ray diffraction patterns of two possible structure types, compared with the estimated (BS – background subtracted) experimental diffraction profile of the $Li_x(C_5H_5N)_yFe_{2-z}Se_2$. Schematics of the layer stacking sequence, when spacers are omitted, in (c) the PbFCl-type structure, adopted by the parent $\beta$-FeSe, and (d) the $ThCr_2Si_2$-type structure, likely adopted by the expanded lattice intercalates.

**Average structure from in-situ perspective.** Ex-situ assessment shows that routinely synthesised $Li_x(C_5H_5N)_yFe_{2-z}Se_2$ yields a relatively small intercalation volume fraction. While the temperature of the solvothermal reaction is an important factor, influence of time as a mediating parameter, together with the molarity (M) of the [Li:PyH5] solution deserve further examination. Time-resolved, in-situ structural studies have demonstrated a unique capability to track the chemical evolution of catalytic[35] and battery[36] materials under "in-operando" conditions, as well as being insightful during the ammonothermal synthesis of alkali[27] and alkaline-earth[28] intercalates of $\beta$-FeSe. Therefore, we designed a total-scattering study to uncover in-situ changes in the average and local structures of the host by varying the molarity (M) of the [Li:PyH5] media, involving freshly made 0.7 M (**A**), 1.4 M (**B**), as well as 72 h aged 0.7 M (**C**) solutions.

The time-variable XRD patterns (Fig. 4a, Fig. S4a,c) in the "Li-PyH5-FeSe" phase-space, do not show any early-time (t< 2 h) Bragg peaks. This confers no intermediate phases, like those emerging upon staging[37] of van der Waals layers in 2D intercalation[38] compounds. However, at somewhat longer times (t≥ 2 h, similar for all molarities), low-q (00*l*) reflections, clearly inform on the formation of the expanded-lattice phase (e.g. q= 0.39 $Å^{-1}$, d= 16.2 Å). The (00*l*) reflections of the $ThCr_2Si_2$ structure type (Fig. S6), do not shift in reciprocal space with time, demonstrating that there are no long-range ordered intermediates between the $\beta$-FeSe and $Li_x(C_5H_5N)_yFe_{2-z}Se_2$ crystallographic phases. This behavior differs from derivatives of alkali[19,27] or alkaline-earth[28] metals with ammonia solutions, where intermediates transform in time with interlayer spacing shrinking (cf. d~ 13.1 Å to 11.5 Å to 9 Å), and where polymorphism, with tunable crystal structures and superconducting properties is feasible.

Indexing of the time-variable XRD patterns (Fig. 4a, Fig. S4a,c), displays (i) growth of intensity at Bragg positions corresponding to the $ThCr_2Si_2$-type [q= 0.39, 0.77 $Å^{-1}$, for (002) and (004) reflections, respectively], (ii) a gradual reduction of scattering assigned to $\delta$-FeSe [hexagonal – $P6_3/mmc$; q~ 2.25 $Å^{-1}$, indexed as (101); Fig. 4a, Fig. S7], and (iii) an irregular behavior of reflections due to parasitic phases of $Li_2Se$ (q~1.8 $Å^{-1}$) and $\alpha$-Fe (q~ 3, 3.5 $Å^{-1}$) (Fig. S3b). Under the chosen conditions the high energy $\delta$-FeSe phase suffers from irreversible decomposition (Fig. S7), supported also by an ex-situ experiment (Fig. S8), but the $\beta$-FeSe phase experiences an equilibrium redox reaction, shifting back and forth with varying [$Li^+/C_5H_5N^-$] concentrations. Moreover, the intensity of the strongest peak characterizing each component phase was normalized to the (101) reflection of *β*-FeSe (ESI §S5) and was plotted against time (Fig. 4b, Fig. S4b,d). The $ThCr_2Si_2$-like intercalated phase yield is comparable at early times (~12-14 h) for all molarities, but saturates at the full time span (~25 h). Taking into account the LeBail extracted relative intensity of phases indexed in experiment (**A**) (Fig. S3b), a volume fraction of ~39(5)% is estimated for the intercalated phase at the ~25 h mark. Notably, the volume fractions of the parasitic phases (e.g. $Li_2Se$) may be maximized before the end of each experiment, and grow when higher Li concentration is used in the solution or even when the solution is aged (Fig. S4b,d).

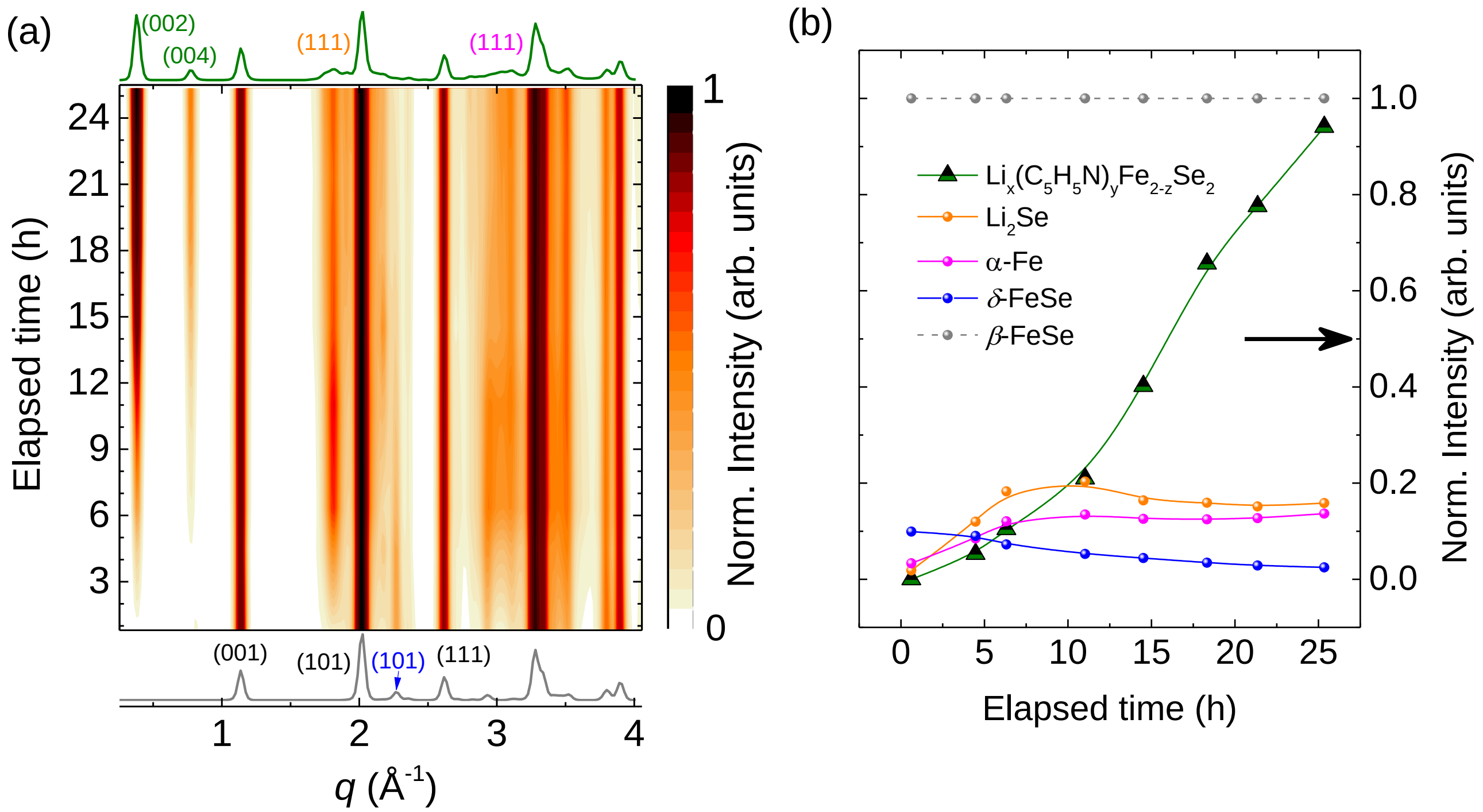

**Figure 4.** Time-resolved evolutions in the Li-PyH5-FeSe phase field (M= 0.7, experiment (**A**). The main Bragg peak in each phase resolved has been normalized to the strongest (101) Bragg peak ($q\cong$ 2.02 $Å^{-1}$) of *β*-FeSe, present at all times (Fig. S5, ESI §S5): (a) 2D film-like plot of the in-situ synchrotron X-ray powder diffraction patterns (λ= 0.1667 Å; sample-to-detector position at 'near' side). The 1D XRD of *β*-FeSe starting material (t~ 0) and the final time-stamp (t~ 25 h) XRD, after the formation of $Li_x(C_5H_5N)_yFe_{2-z}Se_2$, are shown at the bottom and top, respectively. The expanded lattice phase is indexed on the basis of the $ThCr_2Si_2$ structure type, and intense reflections from parasitic phases are also highlighted: $Li_2Se$ ($q$~1.8 $Å^{-1}$; orange), *δ*-FeSe ($q$~2.25 $Å^{-1}$; blue) and *α*-Fe ($q$~ 3, 3.5 $Å^{-1}$; pink). (b) Normalized intensity evolution of the component phases. The strongest reflections utilized, were assigned as: $Li_x(C_5H_5N)_yFe_{2-z}Se_2$, (002) $q$~ 0.39 $Å^{-1}$; $Li_2Se$, (111) $q$~ 1.8 $Å^{-1}$; *δ*-FeSe, (101) $q$~ 2.25 $Å^{-1}$; *α*-Fe, (110) $q$~ 3.10 $Å^{-1}$. A volume fraction of ~39(5)% is estimated for the intercalated phase, at the ~25 h mark.

This demonstrates the critical role of solvated electrons in the reaction pathway. The evolving, strongly reducing character of the solution may instigate adverse effects,[39] including, (a) reduction of FeSe at early times, with a parallel generation of parasitic phases, such as the elemental iron and alkali metal selenides, as for the electrochemical intercalation[40] of $\beta$-FeSe, (b) iron-selenide layer off-stoichiometry, e.g. by excision of Se and/or Fe from the matrix, and (c) a degrading intercalation medium (cf. 72 h aged 0.7 M) involving formation of bulkier species, with smaller diffusion coefficient that may further impede the efficiency of the intercalation process. While lithium is solvated by pyridine, rising concentration may lead to a multitude of intercalating moieties, ranging from solvated entities [cf. $Li^+/C_5H_5N^-$] to radical species that lead to synthesis with issues of reproducibility.

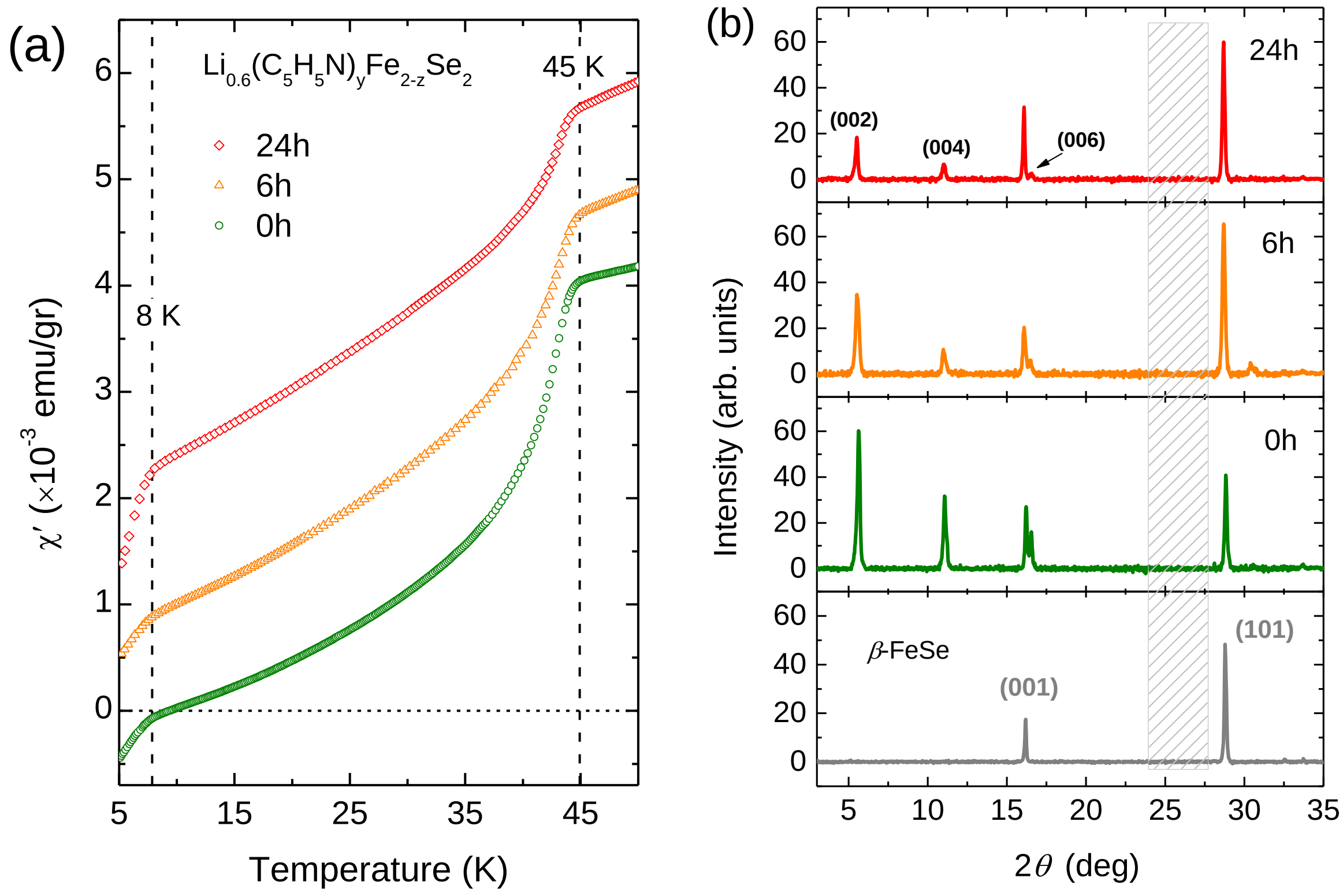

**Figure 5:** Optimizing the synthesis of a formally $Li_{0.6}(C_5H_5N)_yFe_{2-z}Se_2$ sample with respect to the [Li:PyH5] solution aging. Solvothermal reactions with the $\beta$-FeSe were performed, immediately after Li metal dissolution (0 h), with freshly made (6 h) and aged (24 h) liquid intercalation media. (a) AC susceptibility, under $H_{ac}$= 1 Oe, f= 999 Hz, and (b) powder XRDs (CuKa) for the corresponding dry products. Shaded XRD area, 2θ= 24-28°, is an excluded region due to contribution from a home-made, air-tight sample holder.

After identifying the challenges that impact the phase purity, we attempted to further optimize the synthesis in terms of increasing the intercalated phase volume fraction and decreasing the parasitic magnetic moment contribution. We found one promising avenue by reducing the molarity of the [Li:PyH5] solution (M= 0.2) and reacting with $\beta$-FeSe at 80°C. The properties of such ex-situ formed samples were probed by the magnetic susceptibility of the dried products, generated from freshly made (0 h and 6 h) and aged (24 h) [Li:PyH5] solutions (Fig. 5). Here, 0 h means addition of Li metal in PyH5, and immediate reaction of the medium with the host. Comparison of the XRD patterns readily demonstrates that the growth of the expanded lattice phase volume fraction (cf. (002) reflection evolution), is inversely proportional to the solution aging: the fresher the solution the larger the intercalation yield. Importantly, the susceptibility data reveal that the remnant $\beta$-FeSe phase is almost fully consumed (~9% volume) under the chosen solvothermal conditions. Conversely, as the solution becomes progressively aged (24 h), the phase that contributes to the raised paramagnetic moment, grows progressively at the expense of the superconducting phase with the 45 K transition. Reduced molarity combined with use of fresh solution appears to significantly improve the intercalation yield, enhancing the superconducting volume fraction, and simultaneously reducing detrimental magnetic contributions.

**Local structure from in-situ assessments.** In a simple qualitative assessment of the total scattering in the "Li-PyH5-FeSe" phase-field, irregularities in the XRD, I(q) evolution (Fig. 6a-c) map onto PDF, G(r) features (Fig. 6e-g) representing interatomic distances in the component (e.g. from the expanded lattice to $\beta$-FeSe, $\alpha$-Fe, $Li_2Se$) phases. The complexity of the system can be more thoroughly addressed by structure modeling of the PDF data which allows for quantitative assessment of the evolution of local environments (cf. $FeSe_4$ units) in relation to doping.

This challenging task was handled by utilizing multiple crystallographic phases within the structural refinements (ESI §S9). The outcomes for selected time-stamps of the experiment (**A**), where the intercalated phase grows to become the majority over time, are compiled in Figure 7. The analysis from fully converged refinements (Fig. 7a; quality of fit factor, $R_w$~10%), allowed us to extract the partial PDFs (Fig. 7b) for the expanded-lattice $Li_x(C_5H_5N)_yFe_{2-z}Se_2$ phase itself, and compare the time-resolved evolution of the $FeSe_4$ local environment (Fig. 8) against that in the parent $\beta$-FeSe. The extracted parameters, when the Fe-sites retained full stoichiometry, are summarized in Table S1.

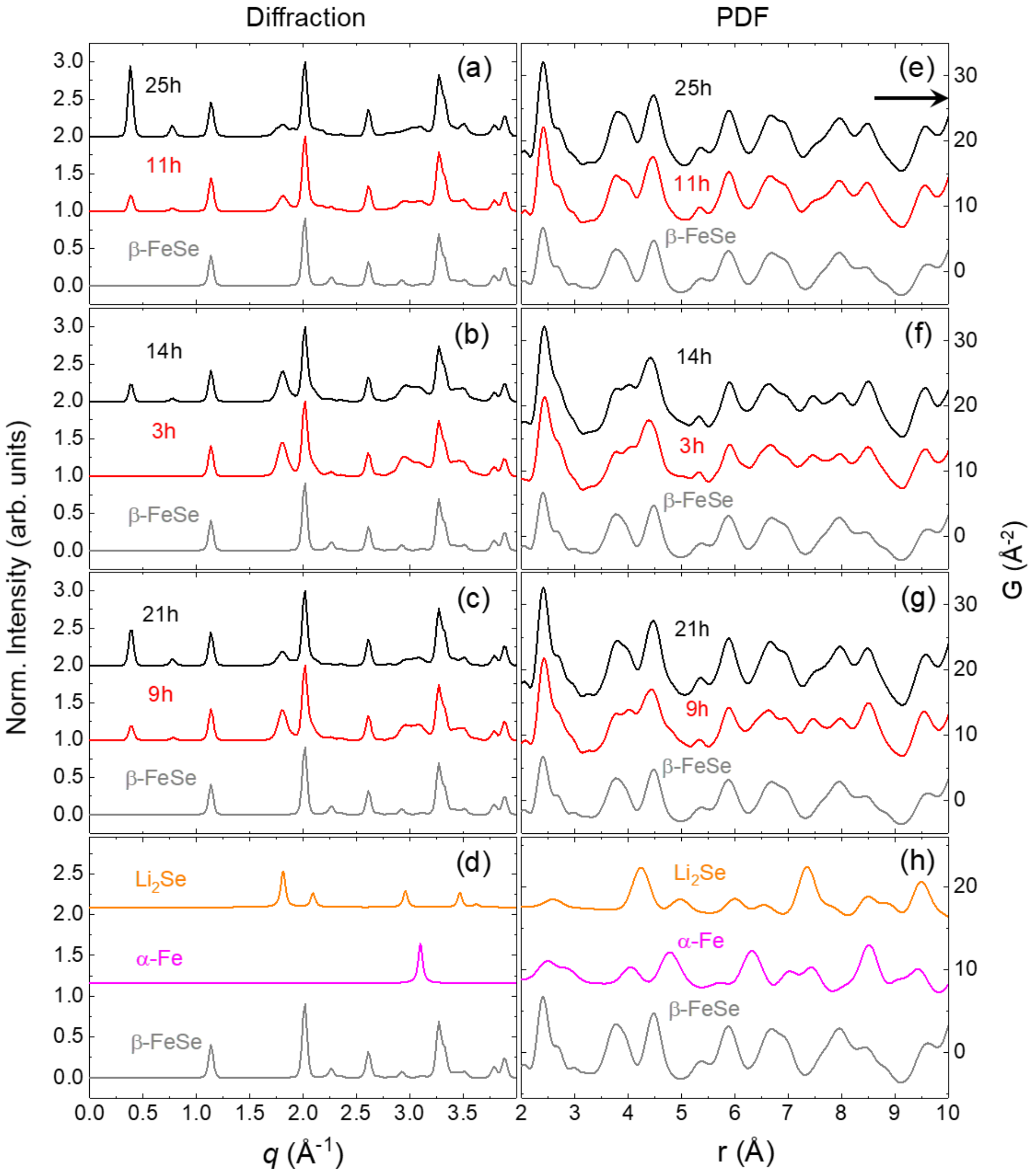

**Figure 6.** Time-resolved, in-situ synchrotron X-ray total scattering data (λ= 0.1667 Å; the sample-to-detector position at 'near' side), for three intercalation experiments in the Li-PyH5-FeSe phase-space, offer simultaneous diffraction, I(q) (a-c) and pair distribution function, G(r) (e-g) views. The $Li_x(C_5H_5N)_yFe_{2-z}Se_2$ phase evolution is mediated by the [Li:PyH5] intercalation medium, assuming, M= 0.7 (a, e), M= 1.4 (b, f), and M= 0.7 aged (c, g), corresponding to experiments (**A**), (**B**) and (**C**). Panels show the starting *β*-FeSe (grey line), the time stamp with parasitic phases in their 'topmost' fraction (red line), and the final time stamp with the in-situ generated material (black line). The simulated powder XRD (d) and PDF (h), corresponding to the major parasitic phases $Li_2Se$ (orange) and *α*-Fe (pink) are also highlighted for comparisons.

separation (d~ 8.6 Å),[21] pinpointing other factors which mediate the electronic properties. Here, while the FeSe slab distance is maintained at d= 16.2 Å, as the reaction propagates, the small elongation of Fe-Se bonds, with respect to $\beta$-FeSe, appears to be invariant with time, but the Fe-Fe distance gradually expands and remains longer (than in $\beta$-FeSe) at the end of the reaction (25 h) (Fig. 8b). Whereas the former may reflect the electron donating nature of the [Li – PyH5] interlayer species, changes in the Fe-Fe distances can involve a complex relation between chemical pressure and iron matrix off-stoichiometry,[13,14] with the last being influenced by the strongly reducing character of the liquid media reaction.

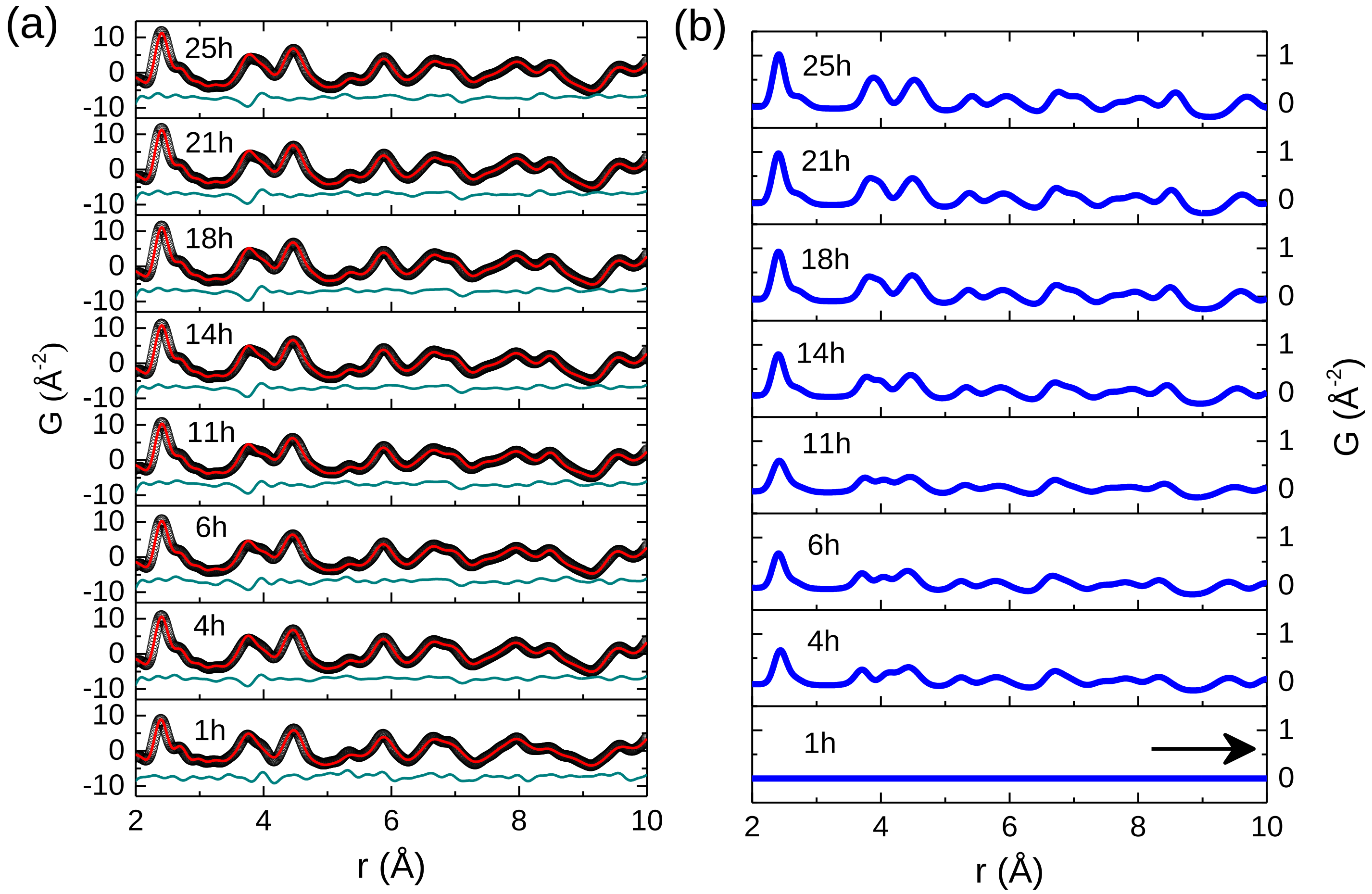

**Figure 7.** Experimental atomic PDFs at variable times during the in-situ structural study of the synthesis of Li/pyridine FeSe intercalates, at 80°C (experiment (**A**)). (a) The G(r) data (black points) are over plotted against the best fit (red line) involving a mixture of crystallographic phases, namely, of $\beta$-FeSe, $\delta$-FeSe, $\alpha$-Fe, and $Li_2Se$, all coexisting with the expanded lattice $Li_x(C_5H_5N)_yFe_{2-z}Se_2$ phase. The difference between model and data, (green line) shown below each time-stamp presented (multiplied by 2 for clarity), is a measure of the quality of the fit ($R_w$~ 10 %). (b) The time-evolution of the derived partial PDFs for the expanded lattice intercalated phase reflect the local structure changes during the insertion of organic molecular spacers in the $\beta$-FeSe matrix.

**Evolution of local distortions in $Li_x(C_5H_5N)_yFe_{2-z}Se_2$.** The discovery of high-$T_c$ (~60 K) superconductivity, with tunable lattice constant in FeSe thin films,[9] has suggested that the electronic structure of Fe-based materials is sensitive to geometric parameters, such as the Fe-Fe and Fe-Se distances,[5] relating its evolution to local environments (cf. $FeSe_4$; Fig. 8a).

For the in-situ $Li_x(C_5H_5N)_yFe_{2-z}Se_2$ phases, as the Fe-Se bond (~2.41 Å) remains almost invariant over time, the expansion in the lattice parameter, a=b= √2× Fe-Fe (Fig. 8b; Table S1), manifests in the variation of the tetrahedral angle, α (Fig. 8c). PDF modeling probes the slow swelling of the correlated $FeSe_4$ units of the matrix ($V_T$, Fig. 8c; Table S1), and suggests that the bond angle increases, with the tetrahedra being highly irregular (α~ 101°), at the beginning of the reaction becoming more compressed (α~ 105°), like in *β*-$Fe_{1+\delta}Se$,[41,42] at the end of the experiment. It is important to understand the underlying reason behind the observed changes. We recall, for example, in the superconducting $K_xFe_{2-y}Se_2$,[43,44] disordered Fe-site vacancies influence the Fe-lattice metrics. Related disorder, probed by synchrotron X-ray absorption spectroscopy, in the expanded lattice $Li_{1-x}Fe_x(OH)Fe_{1-y}Se$ system,[16] correlates local environments and $T_c$. In the latter system, raising the Fe-site occupancy (i.e. [y] smaller), strengthens the Fe-Fe bonding in the FeSe layer, and results in shorter in-plane lattice parameters. Superconductivity emerges with occupancy [1-y]> 95%, when the Fe-square lattice parameter a≤ 3.80 Å ($T_c$= 17 K), while maximal $T_c$ is attained in vacancy-free products (i.e. [y]~ 0), at even shorter a~ 3.75 Å ($T_c$= 41 K).

In the case of $Li_x(C_5H_5N)_yFe_{2-z}Se_2$ (cf. experiment (**A**)), in-plane lattice parameters become progressively longer with time (cf. a=b= √2× Fe-Fe ~ 3.72-3.83 Å) (Fig. 8b, Table S1). Sweeping through the range, one crosses a~ 3.75 Å that in $Li_{1-x}Fe_x(OH)Fe_{1-y}Se$ gives very high $T_c$. The evolution in $Li_x(C_5H_5N)_yFe_{2-z}Se_2$ points to an optimal reaction window of 11-18 h, where in-situ samples, with desirable for superconductivity FeSe slab metrics, a~ 3.75-3.80 Å,[9,16] are formed. Outside this window, the in-situ synthesis generates $Li_x(C_5H_5N)_yFe_{2-z}Se_2$ specimens where the longer the exposure of the starting material with the intercalation medium is, the larger the in-plane lattice constants are. The trend is opposite to that in $Li_{1-x}Fe_x(OH)Fe_{1-y}Se$ where raising of the Fe-site occupancy (i.e. [y] smaller) leads to 'stronger' Fe-Fe bonding. This likely implicates decrease in the Fe-site occupancy in the [Li – PyH5] intercalates at long times, which causes 'weaker' Fe-Fe bonding, leading to stretched Fe square plane. Multi-parametric G(r) fittings, with the occupancy of Fe released in the $ThCr_2Si_2$-type intercalated phase (model differs from that where

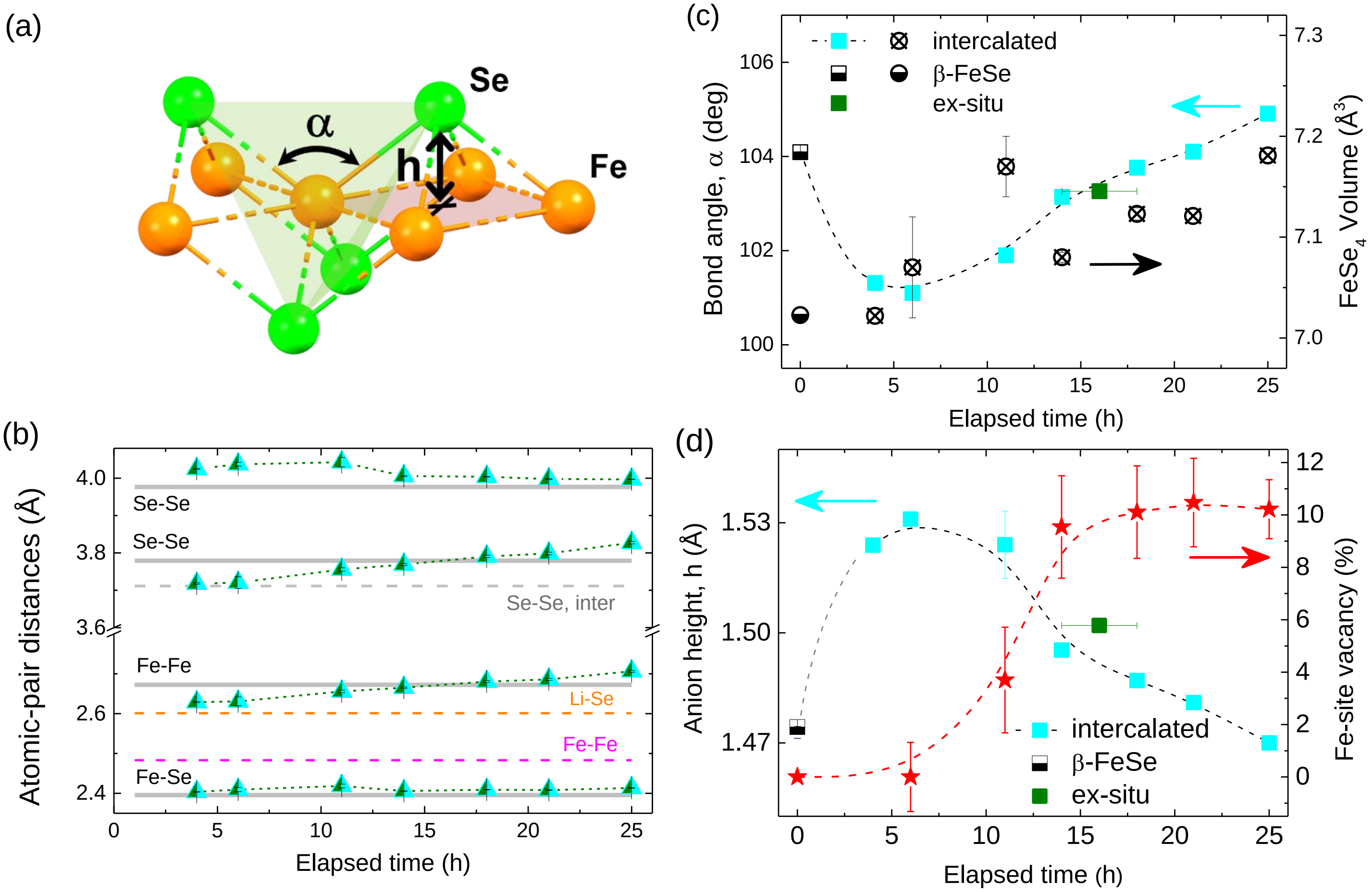

**Figure 8.** (a) Schematic depicting the local-scale, in-plane $FeSe_4$ edge-sharing tetrahedral environment, defined by the Fe-Se bond length and the Fe-Se-Fe angle (α), and its association with likely $T_c$-scaling parameters, such as the Fe square-lattice parameter (cf. Fe-Fe distance) and the anion-height (h). (b) Time-resolved evolution of specific atomic pair distances derived after multiphase fitting of the in-situ obtained G(r) data during the synthesis of Li/pyridine FeSe intercalates, at 80°C (experiment (**A**)). Main atomic radial distances corresponding to the *β*-FeSe starting material (grey) and parasitic phases (*α*-Fe and $Li_2Se$) are presented as pink and orange horizontal lines, respectively. (c) The tetrahedral Se-Fe-Se bond angle (α) and the $FeSe_4$ tetrahedron volume ($V_T$). (d) The distance between Fe and Se planes (i.e. anion height, h), and the Fe-site deficiency (normalized to t= 6 h, where the amount of intercalated phase was detectable for the first time and was vacancy-free). For comparison, in panels (c, d), the parent *β*-FeSe in-plane metrics are drawn at time-zero, while an ex-situ grown $Li_x(C_5H_5N)_yFe_{2-z}Se_2$ (nominal x~2.0) intercalate, is placed near the 14-18 h time stamps.

Fe-site is stoichiometric, i.e. fixed at 1 – Table S1), though with no improved quality of fit, suggest a trend where the Fe-deficiency elevates (~10±1%) towards the end of the reaction (Fig. 8d). The observation that Fe vacancies may be forming at later times could be yet an additional challenge in obtaining the optimal material, emphasizing the important role of Fe-stoichiometry.

The intercalation-induced behavior of the local environments, and their change in relation to doping, can also be correlated with the distance between Fe and Se planes (h, anion height; Fig. 8a), which is considered to be another $T_c$ scaling parameter.[45,46] For example, shrinking such a local parameter,[47] qualitatively reflects on the interplay of electronic and magnetic properties that suppresses superconductivity in $K_xFe_{2-y}Se_{2-z}S_z$.[48] For the in-situ $Li_x(C_5H_5N)_yFe_{2-z}Se_2$, the Fe-Se height progressively shrinks (Fig. 8d) as the [Li – PyH5] interlayer species relax. These samples, at the end of the in-situ experiment, possess an Fe-Se height of about 1.47 Å (Table S1), similar to the value (~1.46 Å) for the lower $T_c$ $K_xFe_{2-y}Se_2$.[49] However, near the optimum reaction window of 11-18 h, 'taller' anion heights emerge, approaching the requirements (h~ 1.52 Å) for high-$T_c$, in expanded lattice materials, made via ammonothermal[19,27] and hydrothermal[50] methods.

Consistent with there being an electron transfer to the host there is a small but discernible elongation of the Fe-Se bonds, which from 2.39 Å in the parent $\beta$-$Fe_{1+\delta}Se$, grow to 2.41 Å in $Li_x(C_5H_5N)_yFe_{2-z}Se_2$ and further up to 2.44 Å, in the case of high-$T_c$ $Li_{0.6}(NH_2)_{0.2}(NH_3)_yFe_2Se_2$. Moreover, insights from the PDF analysis of an ex-situ $Li_x(C_5H_5N)_yFe_{2-z}Se_2$ (nominal x~2.0) material (Fig. S10; Table S1), place such a thermodynamically stable specimen, near the 14-18 h time stamps of the in-situ grown, kinetically stabilized phases (Fig. 8c,d). With prime difference between ex-situ and in-situ (25 h, **(A)**) specimen being the subtle Fe-Se bond elongation (2.42 Å), the PDF analysis suggests that $FeSe_4$ units distort to appear somewhat more swollen, with h~ 1.5 Å in the ex-situ case. Such a span in fundamental local structural properties, indicates that parametrizing the $T_c$, is not a simple relation of the distortion of the shape of the $FeSe_4$ units, but a combined function of the doping level generated by the electron-donating molecules, which could provide optimal superconducting properties for each type of system.

## CONCLUSIONS

The established preparation methods for the Li/pyridine FeSe intercalates have been far from ideal. As such, new approaches to understand the complexity of the intercalation process are sought in order to arrive to improved synthesis protocols for the expanded lattice $Li_x(C_5H_5N)_yFe_{2-z}Se_2$ superconductors. Along this materials' optimization direction, we utilized high-energy synchrotron X-ray total scattering approach, which offers an in-situ view of the structural evolution on distinct length-scales, from intra-unit cell local structure to long-range ordered average structure, during the redox reaction in the Li-PyH5-FeSe phase space.

Despite the large amount of analogous molecule-intercalated $A_x$(amine)$Fe_{2-y}Ch_2$ (A= alkali, alkaline earth; amine= $NH_3$, aliphatic diamines; *Ch*= S, Se) compounds, which react efficiently over wide range of compositions, all the pyridine-derivatives discussed here are mixtures of crystallographic phases regardless of the molarity of the Li metal pyridine solutions or the conditions (t, T) utilized in their preparation. The influence of the strongly reducing Li:PyH5 solution, the aromaticity of the heterocyclic amine and the presence of metastable [$Li^+/C_5H_5N^-$] complexes seem to shift the reaction equilibrium towards parasitic $\alpha$-Fe and $Li_2Se$ phases. Although the intercalation yield of molarities studied (M= 0.7 (**A**), 1.4 (**B**)) is comparable at early times, a lower M solution offers a kinetically improved route, as the $\beta$-FeSe decomposition is less pronounced at all times. Another relevant observation, learned from our in-situ total scattering experiments towards the preparation of the $Li_x(C_5H_5N)_yFe_{2-z}Se_2$ system, is the requirement of a freshly-prepared Li:PyH5 solution, since its aging (**C**) prevailed as intercalation hindrance.

Moreover, we probed the average structure ($ThCr_2Si_2$-type) and interlayer distance (d= 16.2 Å), through which the intercalation yield was estimated by synchrotron powder XRD. In parallel, we also assessed by PDF the local atomic structure distortions involving the fundamental tetrahedral $FeSe_4$ building blocks, which are considered to be scaling parameters for the adjustment of $T_c$. Atomic PDF suggests that as the [Li – PyH5] interlayer species are settled in the matrix, there is a good working time-window where kinetically stabilized phases adopt tetrahedral building blocks that resemble the structural environment required for high-$T_c$ in expanded lattice iron-chalcogenides. The Fe-Se bonds were found a little longer than in the parent $\beta$-FeSe compound, indicating plausible electron donating effects for the $Li_x(C_5H_5N)_yFe_{2-z}Se_2$ intercalates. However, as time elapsed and the intercalation volume fraction grew, the Fe-Fe distances dictated a basal plane expansion, with concomitant shortening of the anion height. Further stretching of the Fe-Fe square planar topology may suggest a delicate interplay of structural defects (cf. Fe-deficient layers) and electronic properties. Such a local structural inhomogeneity in the $Li_x(C_5H_5N)_yFe_{2-z}Se_2$ series with larger in-plane lattice constant, and the ability of interlayer species to raise the doping level, are factors which beyond the optimal FeSe interlayer separation, compete with the superconducting state and mediate the magnitude of $T_c$ through the series.

**ACKNOWLEDGEMENTS**

This research is supported by the Office of Naval Research Global under Award Number N62909-17-1-2126. The work used the beamline 28-ID-1 of the National Synchrotron Light Source II (NSLS-II), a U.S. Department of Energy (DOE) User Facility operated by Brookhaven National Laboratory (BNL). Work in the Condensed Matter Physics and Materials Science Division (CMPMSD) at BNL was supported by the DOE Office of Basic Energy Sciences. Activities at NSLS-II and CMPMSD were supported by the DOE Office of Science under Contract No. DE-SC0012704. We thank Mr George Papadakis for his assistance with the portable reactor for in-situ total scattering studies, designed and manufactured at IESL-FORTH. We are grateful to Mr John Trunk for the help provided during the preparation of the in-situ experiments at NSLS-II.

**SUPPLEMENTAL MATERIAL**

Experiments; Self-fabricated reactor for in-situ total scattering; LeBail refinements; Time-resolved in-situ synchrotron XRDs; Normalization of raw in-situ XRDs; Interconversion of $\delta$-FeSe phase; In-situ atomic PDF refinements; Ex-situ total scattering analysis; Table of local structure parameters; associated with this paper are available free of charge via the Internet at http://.........

**TABLE OF CONTENTS GRAPHIC**

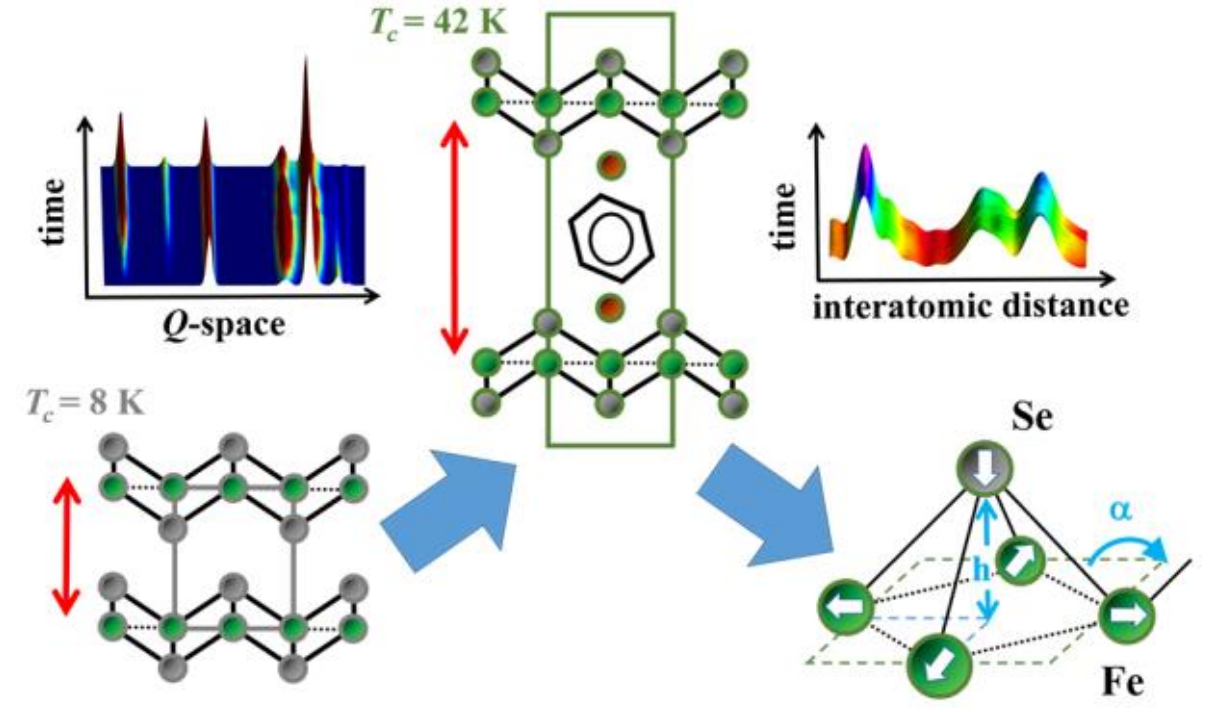

**TABLE OF CONTENTS SYNOPSIS**

Time-resolved synchrotron X-ray total scattering upon intercalation of layered *β*-FeSe by lithium-pyridine moieties reveals enhanced superconducting properties in $Li_x(C_5H_5N)_yFe_{2-z}Se_2$ that are a combined function of stretched in-plane lattice constant and taller anion height, mediated by electron-donating intercalants.

# SUPPLEMENTAL MATERIAL

## In-situ visualization of local distortions in the high-$T_c$ molecule-intercalated $Li_x(C_5H_5N)_yFe_{2-z}Se_2$ superconductor

Izar Capel Berdiell, [1] Edyta Pesko, [2] Elijah Lator, [1] Alexandros Deltsidis, [1,3] Anna Krztoń-Maziopa,[2] A. M. Milinda Abeykoon,[4] Emil S. Bozin,[5] Alexandros Lappas,[1,*]

[1] *Institute of Electronic Structure and Laser, Foundation for Research and Technology–Hellas, Vassilika Vouton, 711 10 Heraklion, Greece.*

[2] *Faculty of Chemistry, Warsaw University of Technology, 00-664 Warsaw, Poland.*

[3] *Department of Materials Science and Technology, University of Crete, Voutes, 71003 Heraklion, Greece.*

[4] *Photon Sciences Division, National Synchrotron Light Source II, Brookhaven National Laboratory, Upton, NY 11973, USA.*

[5] *Condensed Matter Physics and Materials Science Division, Brookhaven National Laboratory, Upton, New York 11973, USA.*

[*] e-mail: lappas@iesl.forth.gr

# Table of Contents

## S1. Experimental Methods

**Ex-situ sample synthesis.** All reagents and subsequent grown sample manipulations, were undertaken inside an Ar-circulating MBRAUN (UNILab) glove box, with less than 1 ppm $O_2$ and $H_2O$ atmosphere. Two reference materials were synthesized, namely: (i) the tetragonal β-FeSe host, was prepared from high purity (at least 4N) elemental iron and selenium by annealing in evacuated and flame-sealed quartz ampoules through a conventional 2-stage elevated temperature method, described by Pomjakushina et al., [1] and (ii) the hexagonal *δ*-FeSe polymorph, was made by solid-state reaction, where Fe (7 mmol) was mixed with Se (7.7 mmol) and pressed in pellet form (∅ 4 mm). The *δ*-FeSe phase was flame-sealed in quartz ampoule under vacuum (~$10^{-2}$ bar) and reacted in a preheated oven at 775°C for a period of 48 h, when it was quenched in a water bath. $Li_x(C_5H_5N)_yFe_{2-z}Se_2$ intercalates were prepared from 9 mmol of metallic Li (99.9%, Sigma-Aldrich) and 9 mmol of β-FeSe that were packed into one of the compartments of a Schlenk vessel under argon. 15 ml of anhydrous pyridine (99.8%, Sigma-Aldrich) or deuterated pyridine (99.96% DEUTERO, GmbH) was initially put into the second compartment of the Schlenk vessel. At that time pyridine was transferred isopiesticly into the Li and β-FeSe containing section. "Isopiestic" transfer relies on the transportation of the solvent (between two compartments of an evacuated to low pressure Schlenk vessel) that has previously been solidified at the temperature of liquid $N_2$ and pumped down to a pressure of ~$10^{-3}$ Torr. In this operation, the solidified solvent ($C_5H_5N$) is slowly heated up under static vacuum (~$10^{-3}$ Torr) in one of the chambers of the Schlenk vessel, while its vapors condense in the other chamber, which is simultaneously cooled down under liquid $N_2$. The reactants were then brought to room temperature and the reaction took place at 80°C (with the use of an oil-bath) for a period of 24 h. At the end of the time, the excess of the solution was decanted to the second compartment of the vessel and the solid residue was isopiesticly washed with $C_5H_5N$ and then dried under vacuum for 30 min.

**Powder X-ray diffraction.** All the operations on the polycrystalline matrices were performed under argon atmosphere in order to avoid oxidation. Phase purity of the prepared materials was tested with an X-ray powder diffractometer (BRUKER D8 ADVANCE), with Cu-$K_\alpha$ radiation ($\lambda$ = 1.5418 Å), using a dedicated home-made, air-tight sample holder designed for measurements of environmentally sensitive samples.

**Total-scattering experiments.** Synchrotron X-ray total scattering measurements were carried out at the 28-ID-1 beamline at the National Synchrotron Light Souce-II of Brookhaven National Laboratory. Rapid data acquisition mode, with a monochromatic 74.375 keV (λ = 0.1667 Å) X-ray beam, allowed for patterns of a calibrant (Ni) and the highly crystalline reference material (β-FeSe) to be collected in just 5 s, while for the intercalated sample derivatives data acquisition time was 60 s. For such purposes, the fast-responding two-dimensional (2D) PerkinElmer area detector was used. The station's highly versatile setup, designed to allow for rapid changing between chosen positions of the fast-responding area detector, facilitates switching between complementary experimental modes. In this work, two different sample-to-detector positions were used, namely 326 mm (near) and 1124 mm (far) calibrated against the Ni-standard that enable, either broad q-range appropriate for PDF data collection mode or instead higher-angular resolution, for X-ray powder diffraction (XRD) data measurements. Data integration was performed with FIT2D [2,3] software. 2D images were binned and summed together to provide the equivalent of longer exposures for quantitative data analysis. The reduction of I(q) data and their Fourier transformation into G(r) was done using the PDFgetx3 package, while the pair distribution function analysis was carried out with PDFgui [4] suite of programs. LeBail type of full profile analysis of the powder XRD patterns was performed with the FullProf [5] suite.

For the in-situ X-ray total scattering study of the $Li_x(C_5H_5N)_yFe_{2-z}Se_2$ phase formation, we developed a portable reactor setup, which as an integral part of the 28-ID-1 experimental station, allowed us to probe the real-time reactivity of β-FeSe layers in liquid-media (Fig. 1). This home-made reactor, was designed to contain the Li-PyH5-FeSe reactants inside a quartz tube (O.D. 3 mm, 10 cm long), capable to maintain anaerobic conditions while the sample was rotated (Fig. S1). As the materials' growth is regulated by varying the molarity (M) of the [Li:PyH5] solution, within the set time-frame of the synchrotron facility study, three in-situ experiments assuming, freshly made 0.7 M (**A**), 1.4 M (**B**), as well as 72 h aged 0.7 M (**C**) cases, were deployed to probe the synthesis of $Li_x(C_5H_5N)_yFe_{2-z}Se_2$ at 80°C. Effectively, 28-ID-1, with its wavelength tunability and high-flux, is ideal for in-situ work as it can afford views from fast PDF combined with powder XRD, even from a small sample volume of an ever evolving specimen. The sample temperature was controlled with the use of $N_2$ cryostream. The time-resolved processes were then studied not only with sufficient r-space resolution ($q_{max}$~24 $Å^{-1}$) to explore local structure modifications (e.g. due to specimen doping), but also with adequate q-space resolution to study the longer length scales (e.g. interlayer separation during specimen intercalation). Thus, the reaction was sampled at a relatively high rate, with the final time-resolution tuned by digital manipulation of the images (FIT2D).

**Magnetic measurements.** Magnetic measurements were obtained by utilizing an Oxford Instruments MagLab EXA 2000 vibrating sample magnetometer. AC susceptibility as a function of temperature was measured in an applied field of 1 Oe and frequency of 999 kHz. Samples were contained in Quartz ampules (O.D. 5 mm, wall thickness 1 mm). These were filled with polycrystalline β-FeSe (38 mg) and stirred [Li:$C_5H_5N$] solution (0.7 M, 0.25 ml), then closed with parafilm, an NMR cap and Kapton tape to avoid loss of evaporated pyridine (Fig. S2). The ampules were heated in a sand bath at 80°C inside the glove box and while they were intermittently shaken, at given times (8 h, 18 h and 28 h), one ampoule at a time, was removed from the glove-box. It was then flame sealed and measured directly in the magnetometer inside the mother liquid. At completion of the data acquisitions, ampoules containing wet samples, were opened inside the glove box and dried with hexane prior to XRD collection and subsequent characterization back again in the magnetometer.

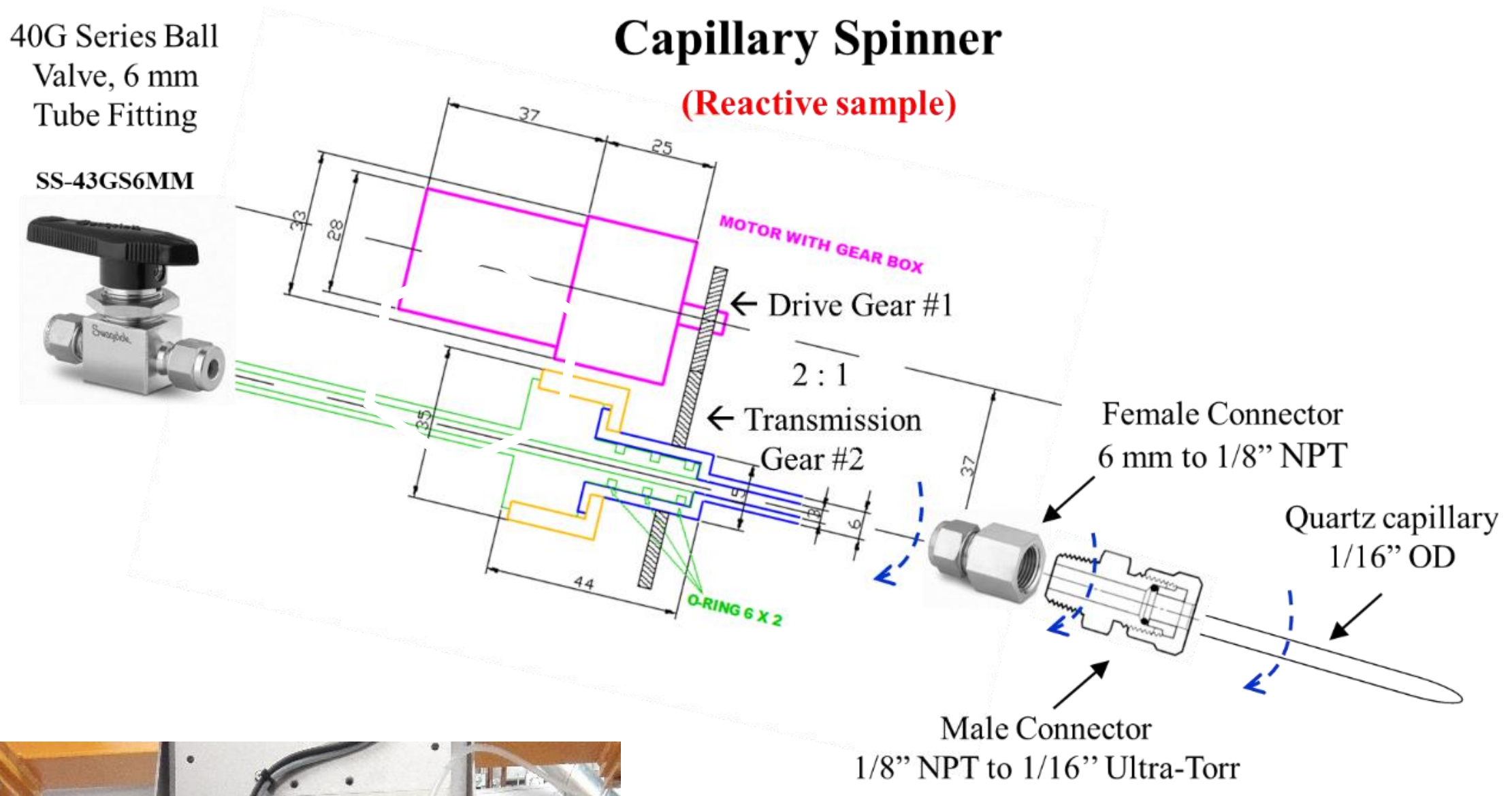

(a)

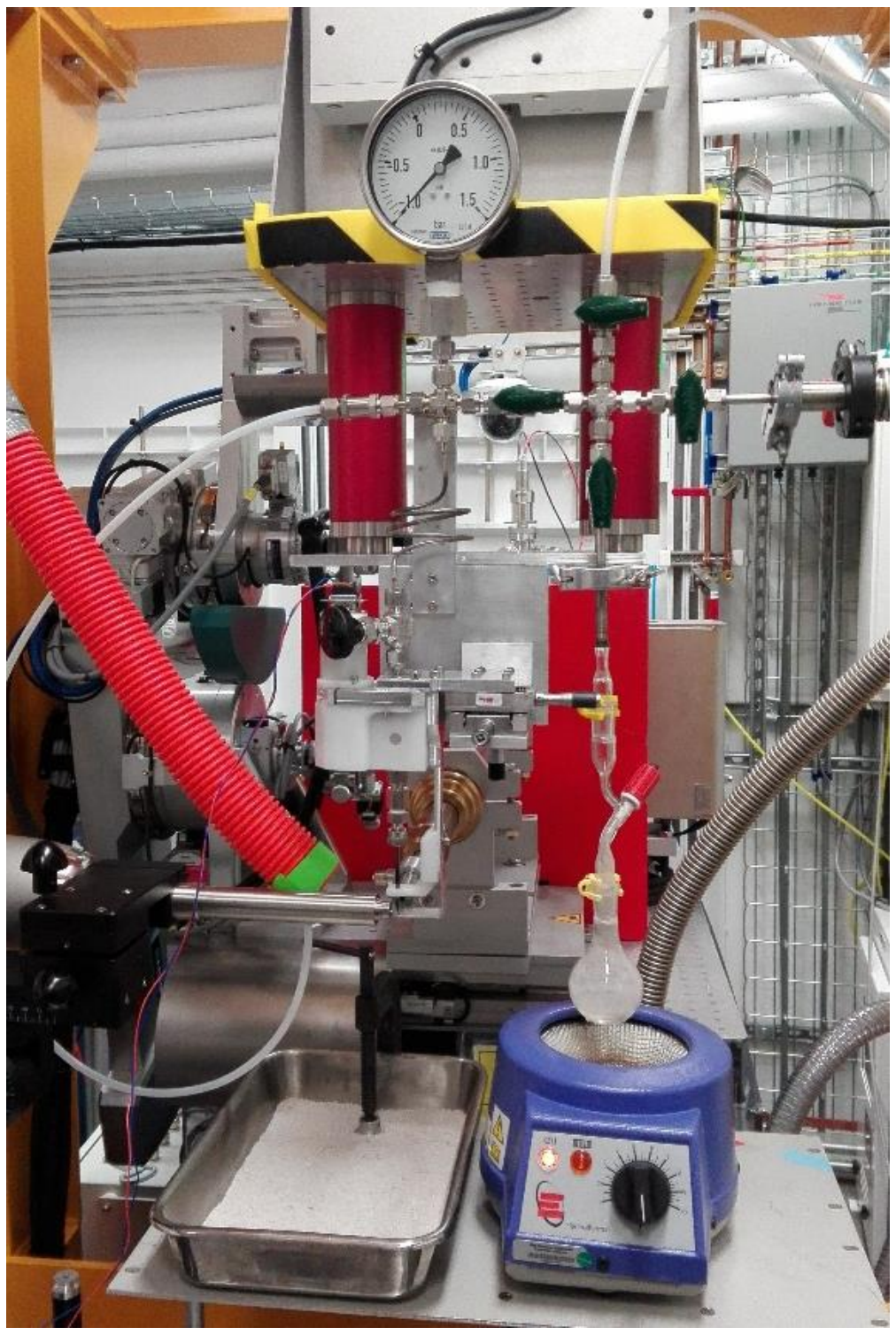

(b)

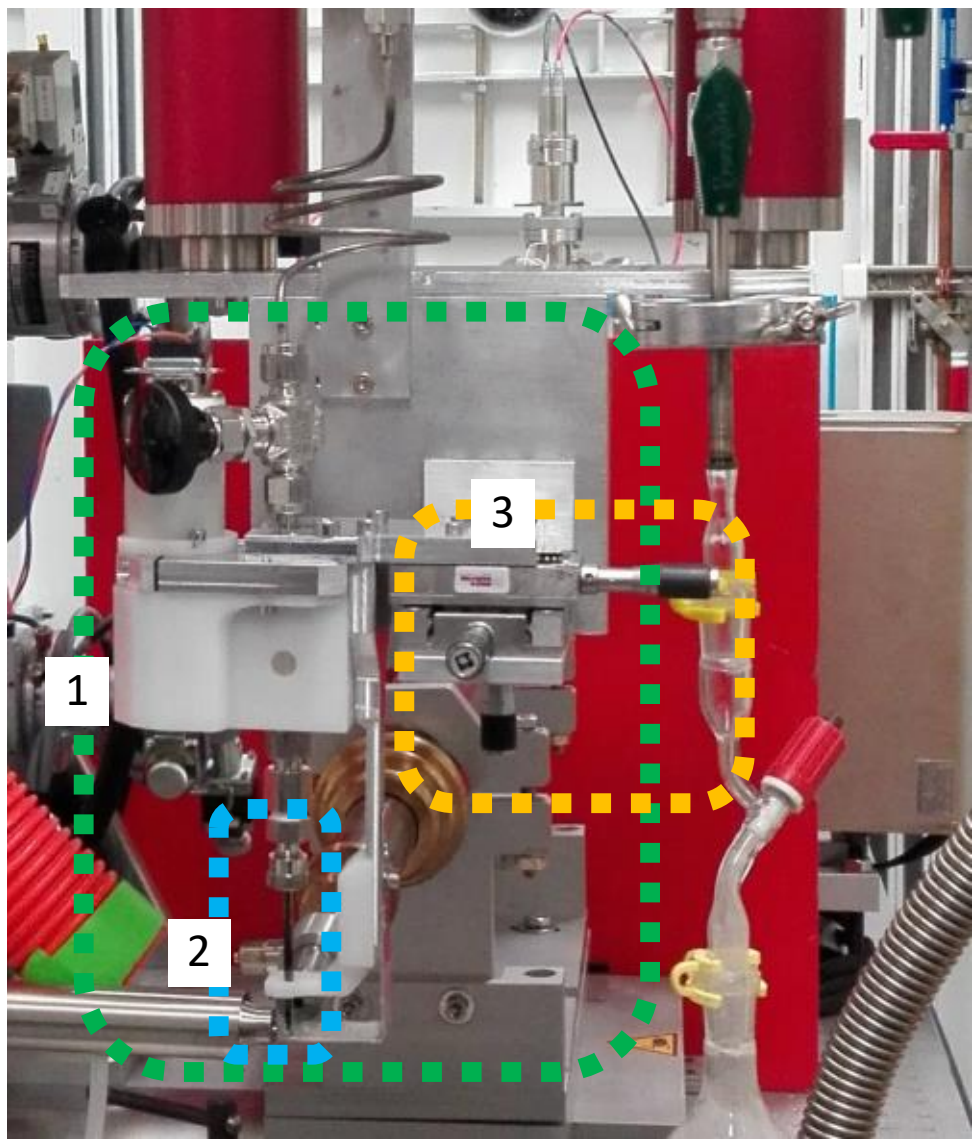

(c)

Figure S1. A portable reactor setup for in-situ total scattering: (a) schematic of the compact air-tight sample rotation module (rotating shaft in blue, powered by a 24 V DC motor via the gear no. #2) spins the reactive sample (at 10-40 turns/min) contained in a thin-wall quartz tube or capillary. The module can be safely disconnected and transferred to an anaerobic glove box for further manipulation of the air-sensitive sample. Dimensions - mm. (b) Photos of the self-fabricated reactor setup, spinning a 10 cm long quartz tube (O.D. 3 mm) installed at the optical table of the 28-ID-1 beam line. (c) Magnified section: the module (1- green-rectangle) that can be mounted and dismounted from the beam line with a Swagelok isolating needle valve; the quartz tube (2- blue coloured rectangle), for the reactants and the x-y-z stage (3- white orange) for adjusting the sample position in the X-ray beam pathway.

## S1.1 In-situ sample synthesis

Procedure for the in-situ synchrotron X-ray total scattering experiments utilizing a home-made portable reactor setup (Fig. S1a) – peripheral components are identified in Fig. 1 (i to v), in main manuscript.

(a) reactor (i) is loaded with β-FeSe in a glove box, then connected to the setup (Fig. S1b-c) at 28-ID-1

(b) reactor is evacuated (~$10^{-4}$ mbar) (ii) and purged with He (iii), then cooled to T~ 0°C (iv)

(c) PyH5 kept at R.T. (v), is allowed to condense on the reactants (i), by the "isopiestic technique"

(d) pyridine (v) source is isolated, and reactor (i) is warmed to R.T. in a controlled way

(e) powder sample in liquid media (i) is warmed to 40 or 80°C (iv) – PyH5 vapor pressure ~60 or 320 mbar, respectively

(f) collect time-dependent patterns from solution of reactants over a period of 8-25 h

(e) with no changes to patterns, PyH5 supply (v) is cooled down to T~ – 5 °C (PyH5 vapor pressure ~4.5 mbar)

(h) excess PyH5 of reactor (i) is re-condensed back to the storage container (v) – the latter is isolated

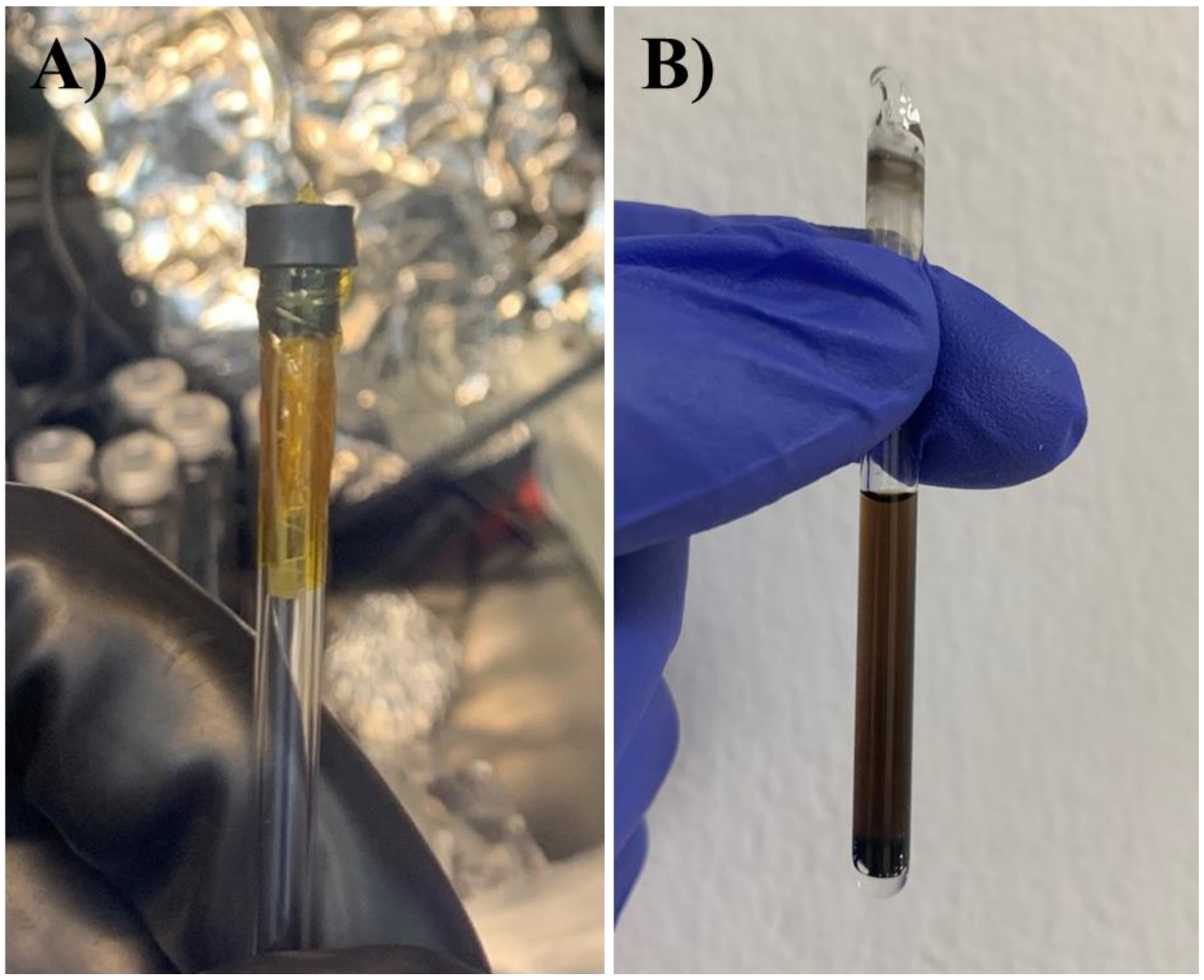

Figure S2. Photos of the quartz ampoules used for the AC magnetic susceptibility experiments. Example of ampule (A) covered with parafilm layer, closed with NMR plastic cap and sealed with Kapton tape, which showed no solvent evaporation after 7 days at 80°C; (B) flame-sealed ampoule, with "Li-PyH5-FeSe reactants in the mother liquid, used for magnetic measurement at variable times (i.e. 8 h, 18 h and 28 h).

## S3. Average structure of ex-situ grown samples by synchrotron XRD

In view of the complex behavior encountered with low-T solvothermal reactions (40-80°C, over 24 hr), we employed the brilliance of synchrotron X-ray powder diffraction to study the room-temperature average structure of a dried $Li_x(C_5H_5N)_yFe_{2-z}Se_2$ (nominal x~2.0) material, prepared in advance. Here, the area detector was positioned at the 'far-side' with the respect to the sample location. Despite the medium resolution of the setup, identification of the different crystallographic phases was possible well-beyond what lab-based CuKa powder diffraction could provide. Indexing of the synchrotron powder XRD (Fig. 3a) and LeBail full profile analysis (Fig. S3a), indicate that such typical, ex-situ grown sample entails multiple component phases, namely, $\alpha$-Fe, $Li_2Se$, and β-FeSe (tetragonal – P4/nmm). All these coexist with a significant amount of the intercalated, expanded lattice 42 K superconducting phase, implying a sub-optimal synthesis procedure, with issues on reproducibility.

Although the parasitic phases present many overlapping reflections, the $Li_x(C_5H_5N)_yFe_{2-z}Se_2$ component phase in the pattern allows predicting the basic structure of this expanded-lattice Fe-chalcogenide, [6] with an interlayer FeSe distance of d= 16.2 Å, determined by the low-angle (00l) Bragg reflection (q= 0.39 $Å^{-1}$, $2\theta$= 0.55°, $\lambda$= 0.1667 Å). Fortuitously, since interlayer species (guests) are not well visible by X-rays, this allows one to focus on the FeSe framework without complications due to the lightly scattering components. What we need to consider here is that in the Fe-based superconductors the common structural building feature is a layer of Fe-chalcogen tetrahedral units. However, adjacent layers, though separated by electron-donating spacers, can be either co-aligned out-of-plane or alternating in orientation along the c-axis. Either type of layer stacking sequence, if spacers are omitted, is present in exemplary superconducting systems, such as the LiFeAs [7] (PbFCl-type) [8] and $BaFe_2Se_2$ [9] (or $ThCr_2Si_2$-type) [10] compounds that are classified as 111 and 122 type (Fig. 3c and 3d), respectively. In view of these topologies, we simulated the powder patterns of 111 and 122 types of FeSe frameworks and compared them with the observed experimental data (Fig. 3b). The latter was an approximation as it was derived by mathematically subtracting the normalized β-FeSe synchrotron X-ray powder pattern, measured under the same exact experimental conditions, while the $Li_2Se$ and α-Fe (colored in orange and pink, respectively; Fig. 3a) pattern regions were manually excluded. This qualitative comparison of the simulated and the measured experimental diffraction profiles, suggests that the $Li_x(C_5H_5N)_yFe_{2-z}Se_2$ may be indexed on the basis of the I4/mmm symmetry (with cell size, a=b≅ 3.8 Å , c≅ 32.47 Å) met in the $ThCr_2Si_2$ structure type.

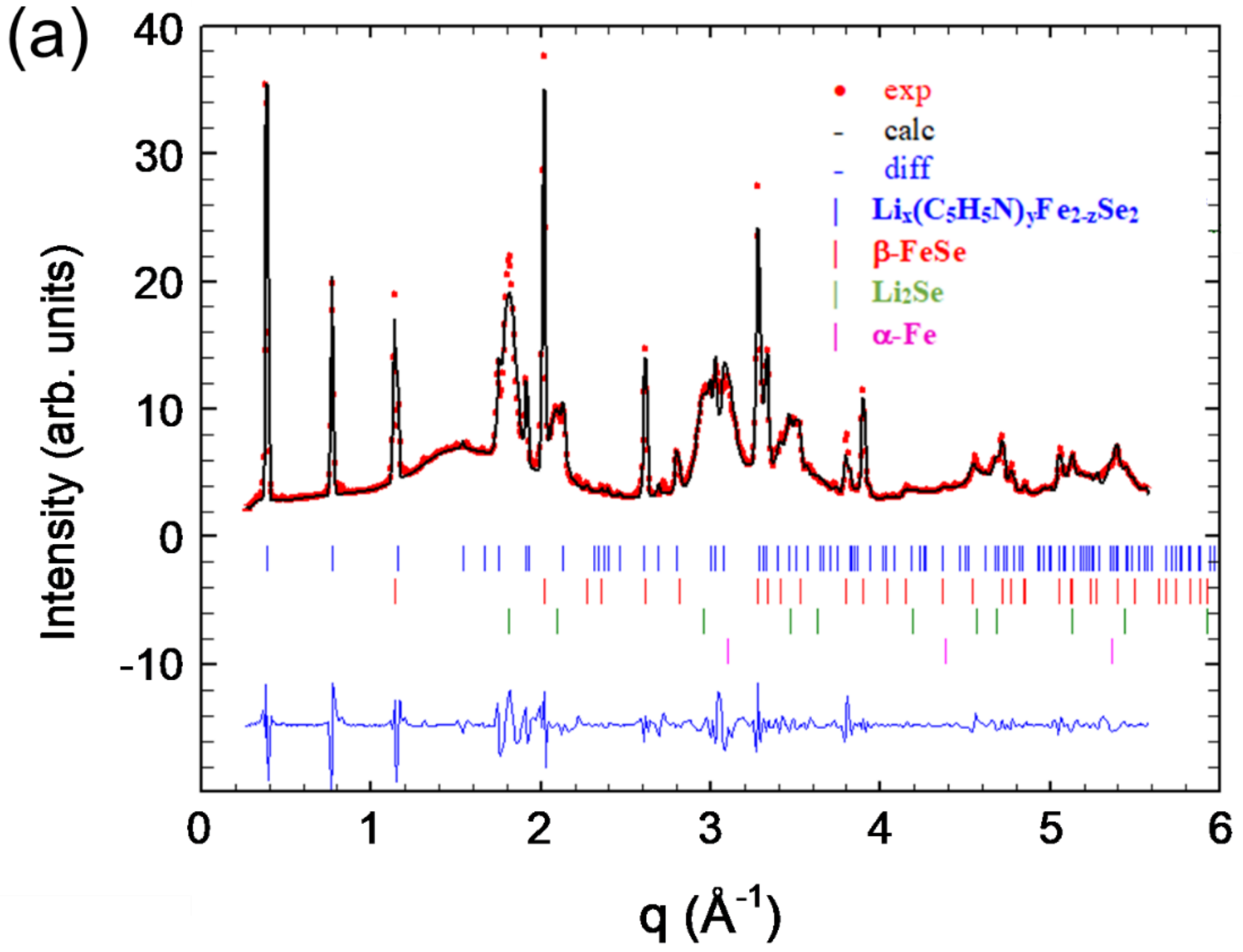

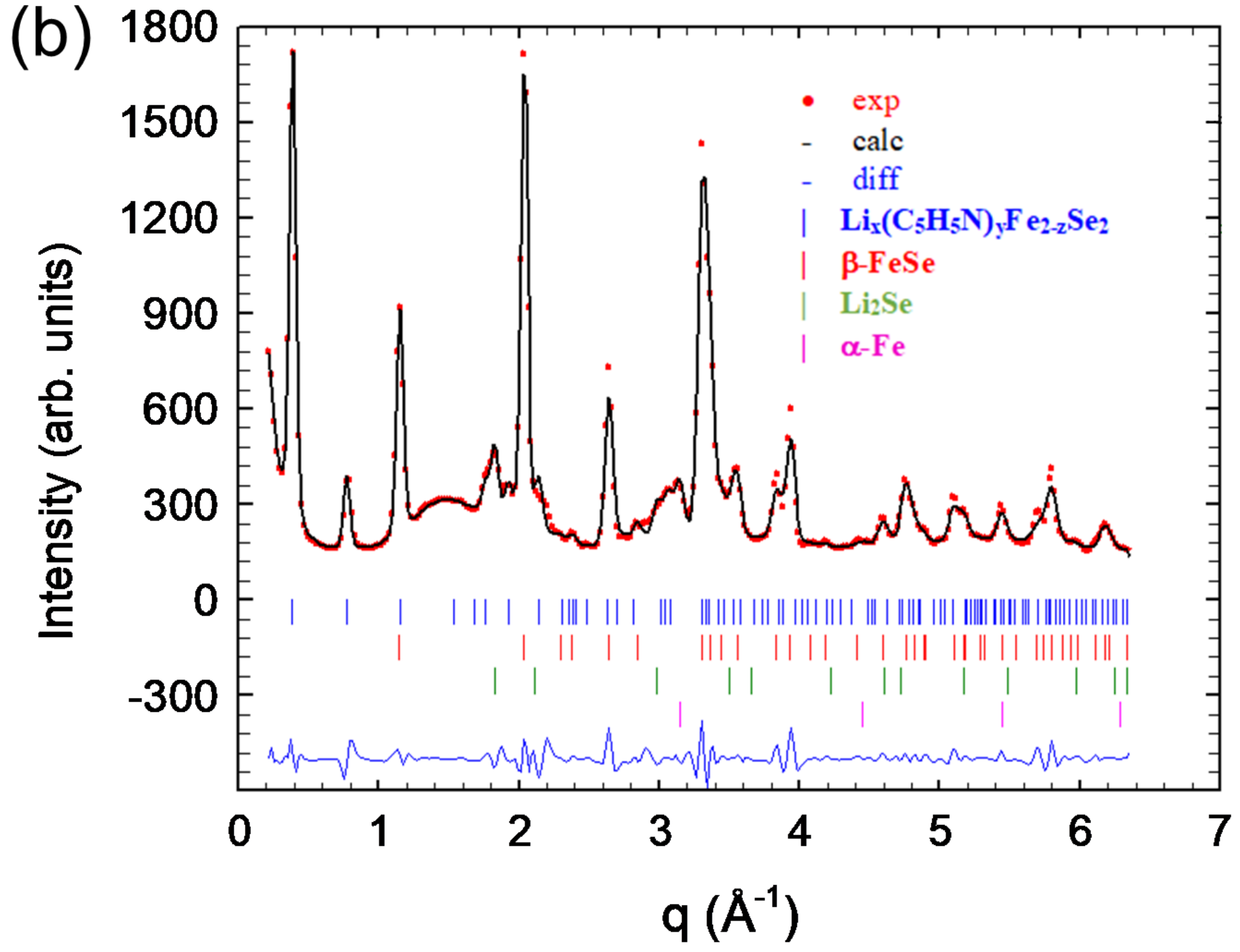

Figure S3. LeBail full profile analysis of synchrotron XRDs: (a) typical, ex-situ grown sample, measured with detector-to-sample at ‘far’ side, (b) an in-situ grown sample (0.7 M, experiment (**A**)), obtained at the end of the experiment (25 h), having the detector-to-sample at ‘near’ side. Multiple secondary phases coexist with the expanded lattice phase, which is indexed on the basis of the $ThCr_2Si_2$ structure type (I4/mmm symmetry; a=b≅ 3.8 Å, c≅ 32.47 Å).

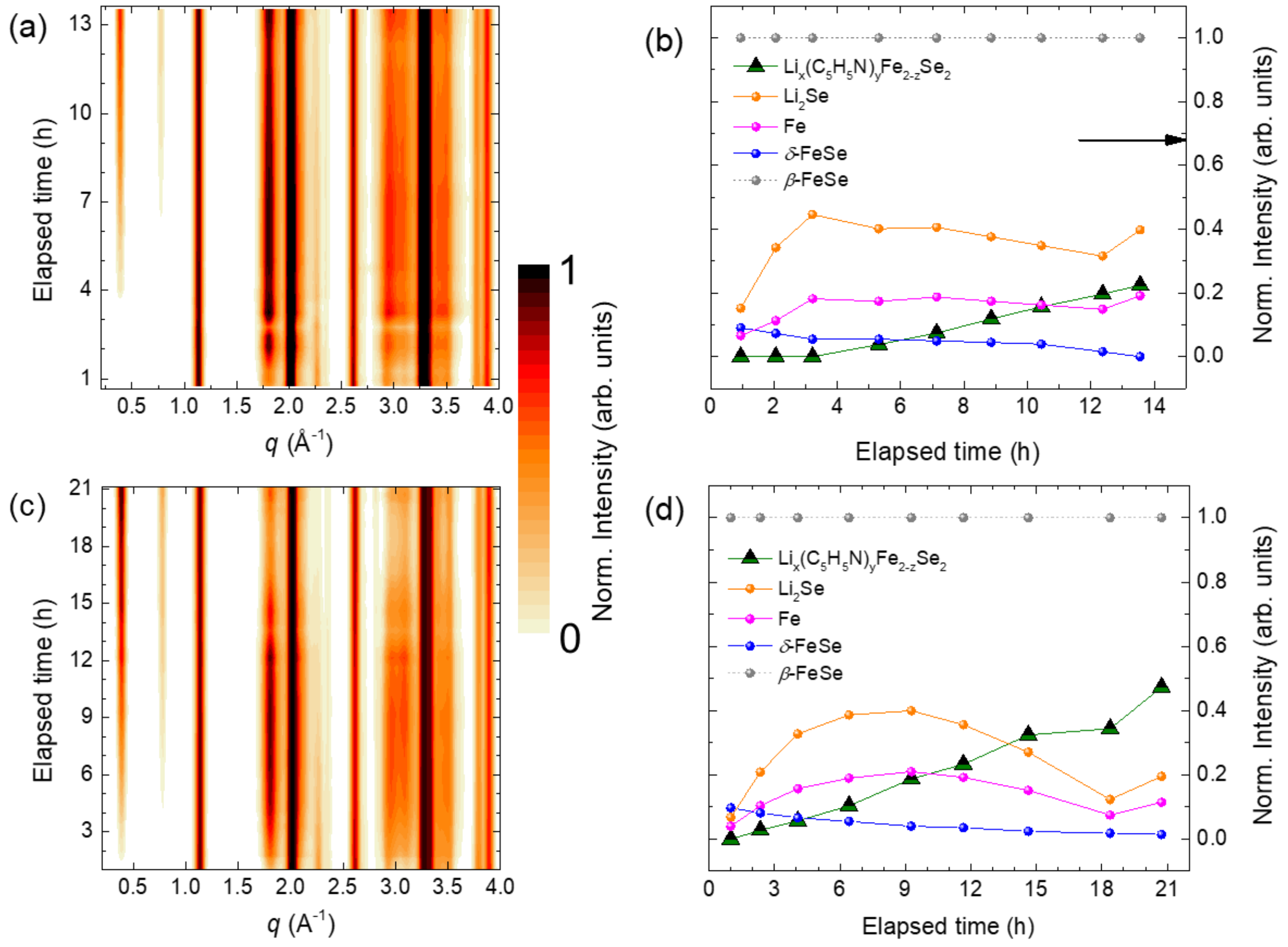

Figure S4. Time-resolved in-situ synchrotron X-ray powder diffraction patterns (λ= 0.1667 Å; the sample-to-detector position at 'near' side) obtained by varying the molarity (M) of the [Li:PyH5] solution (a, c). The evolution of the normalized intensities, (b, d) for the main reflections in each crystallographic phase observed at the average level during the intercalation, namely: $Li_x(C_5H_5N)_yFe_{2-z}Se_2$, (002) $q$ = 0.39 Å$^{-1}$; $Li_2Se$, (111) $q$ = 1.81 Å$^{-1}$; β-FeSe, (101) $q$ = 2.02 Å$^{-1}$; $\delta$-FeSe, (101) $q$ = 2.26 Å$^{-1}$; $\alpha$-Fe, (110) $q$ = 3.10 Å$^{-1}$. Molarities assessed within the set time-frame of the synchrotron facility study were, of 1.4 M (experiment (**B**)) (a, b), and of 72 h aged 0.7 M (experiment (**C**)) (c, d).

## S5. In-situ synchrotron XRD raw data normalization

Raw intensities of the whole diffraction pattern during the in-situ studies varied significantly between different time-stamps within the same molarity data collection (Fig. S5). This reflects the intrinsic character of the experiment where a spinning reactor introduces an additional variable, namely, the amount of sample hit by the incident beam. Therefore, individual background treatment was necessary, i.e. raw data was plotted and the baseline for each time-stamp was carefully selected point by point, interpolating it with a line, and finally was subtracted. Moreover, post normalization was carried out with respect to the starting β-FeSe material [i.e. its strongest (101) reflection, at q≅ 2.02 $Å^{-1}$], present throughout the times, which was acting as an internal calibrant for the overall scale factor. Corrections were applied to every selected time-stamp for interpretation, and systematic assessment of the reaction evolution as function of time.

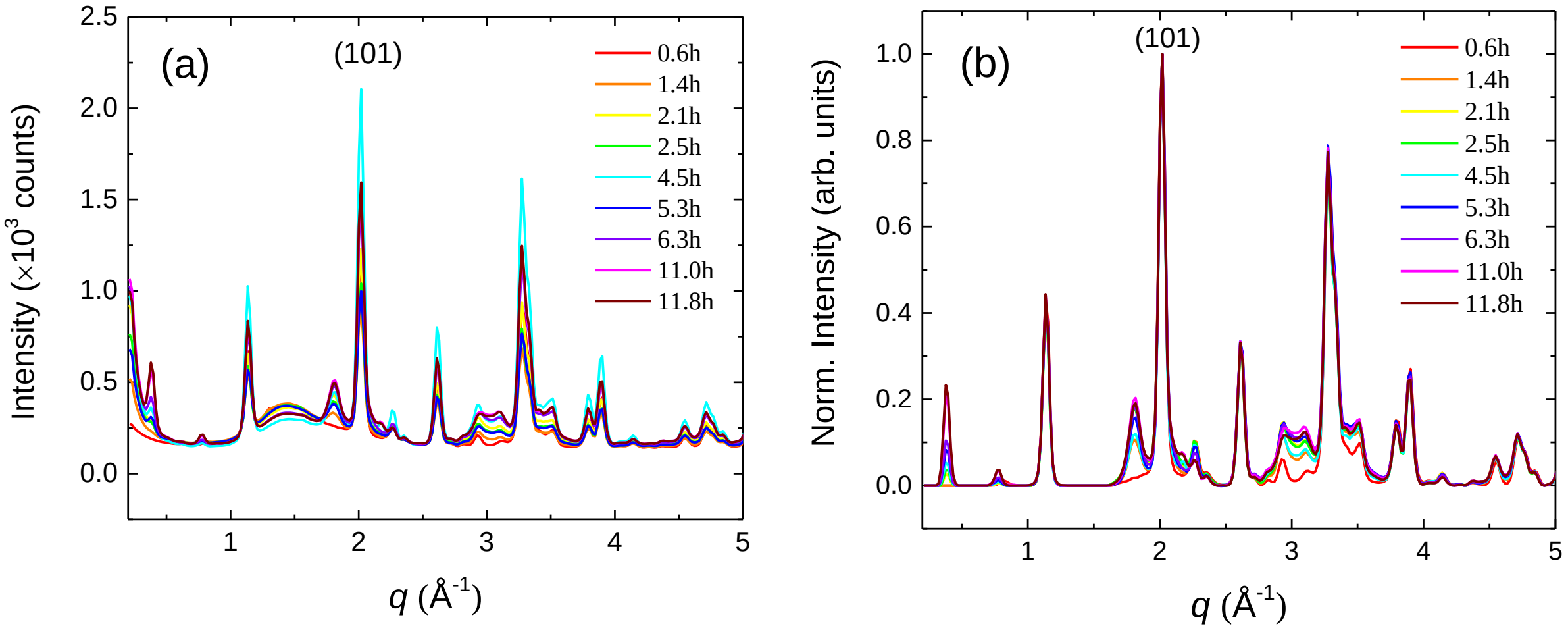

**Figure S5. Variation of in-situ synchrotron X-ray diffraction patterns** (λ= 0.1667 Å) in the "Li-PyH5-FeSe" phase-space, for early time stamps of experiment (**A**) (freshly made 0.7 M), marked in hours: (a) raw data, where the overall scale factor is randomly varying, (b) patterns normalized to (101) reflection of β-FeSe, after manually processing the data to correct for background variation over time.

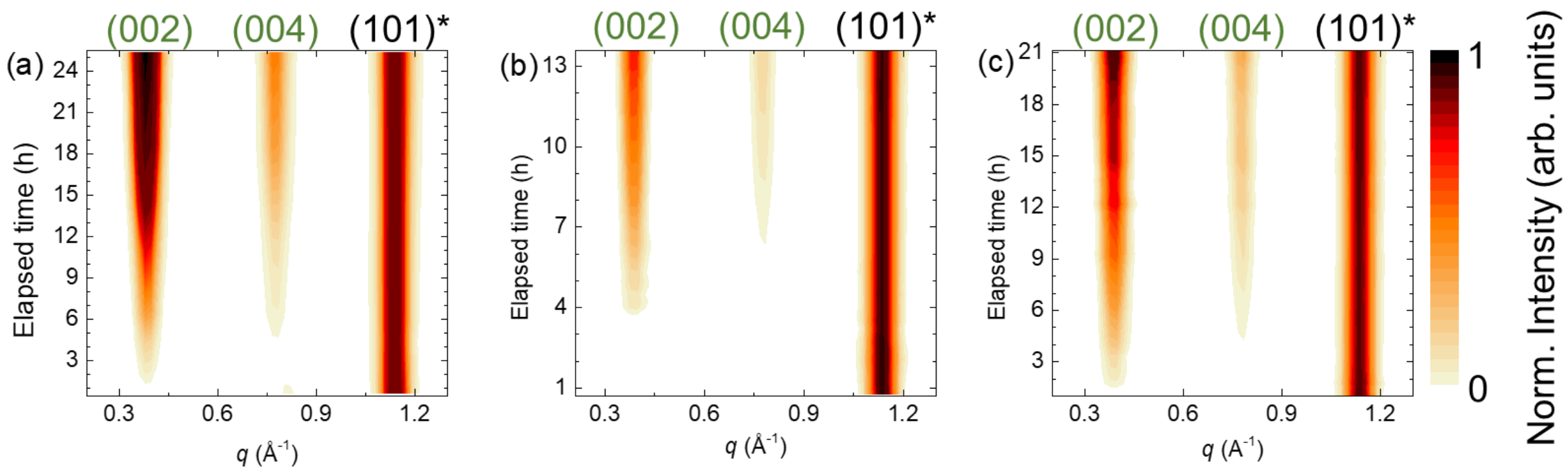

Figure S6. Low-q region of the in-situ synchrotron XRD patterns (λ= 0.1667 Å) for three in-situ experiments (cf. of varying molarity (**A**), (**B**), (**C**)), where the absence of staging is suggested by the lack of appearance and disappearance of early-time Bragg peaks (reflections (002) and (004)). The most intense reflection from β-FeSe starting material, is also shown with index (101)*, in grey color. The 'end-time' for each in-situ experiment is different as it corresponds to variable overall reaction time in each case.

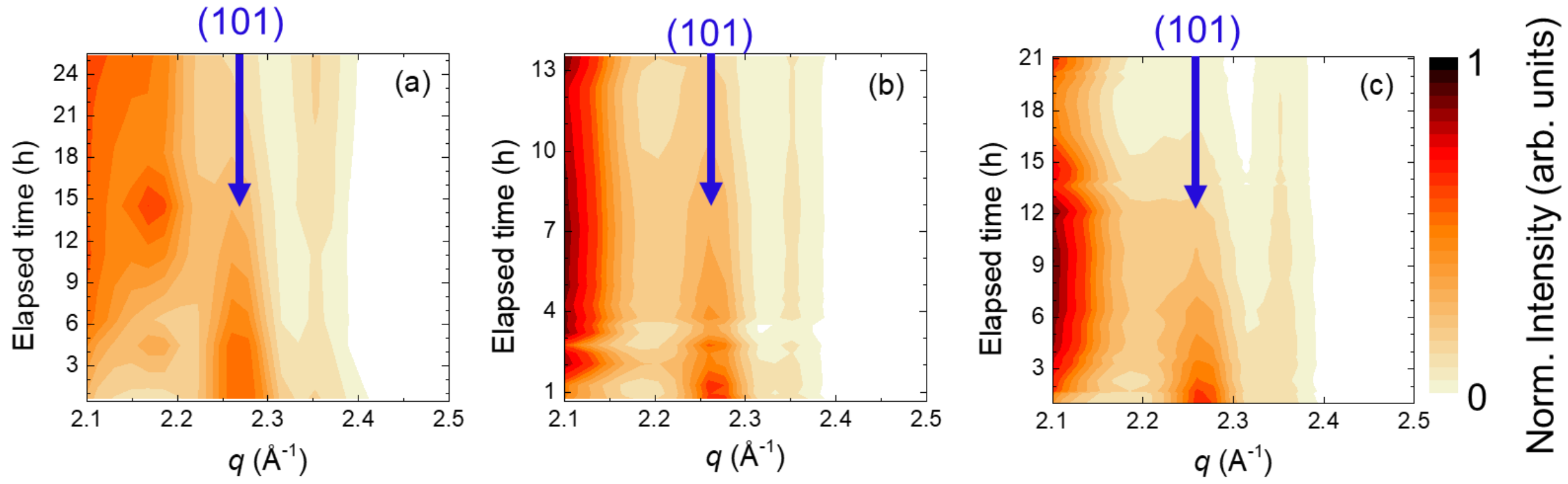

Figure S7. Higher-q in-situ synchrotron XRD patterns (λ= 0.1667 Å) for three in-situ experiments (cf. of varying molarity (**A**), (**B**), (**C**)), where the progressive disappearance of (101) reflection, belonging to hexagonal *δ*-FeSe phase, is highlighted.

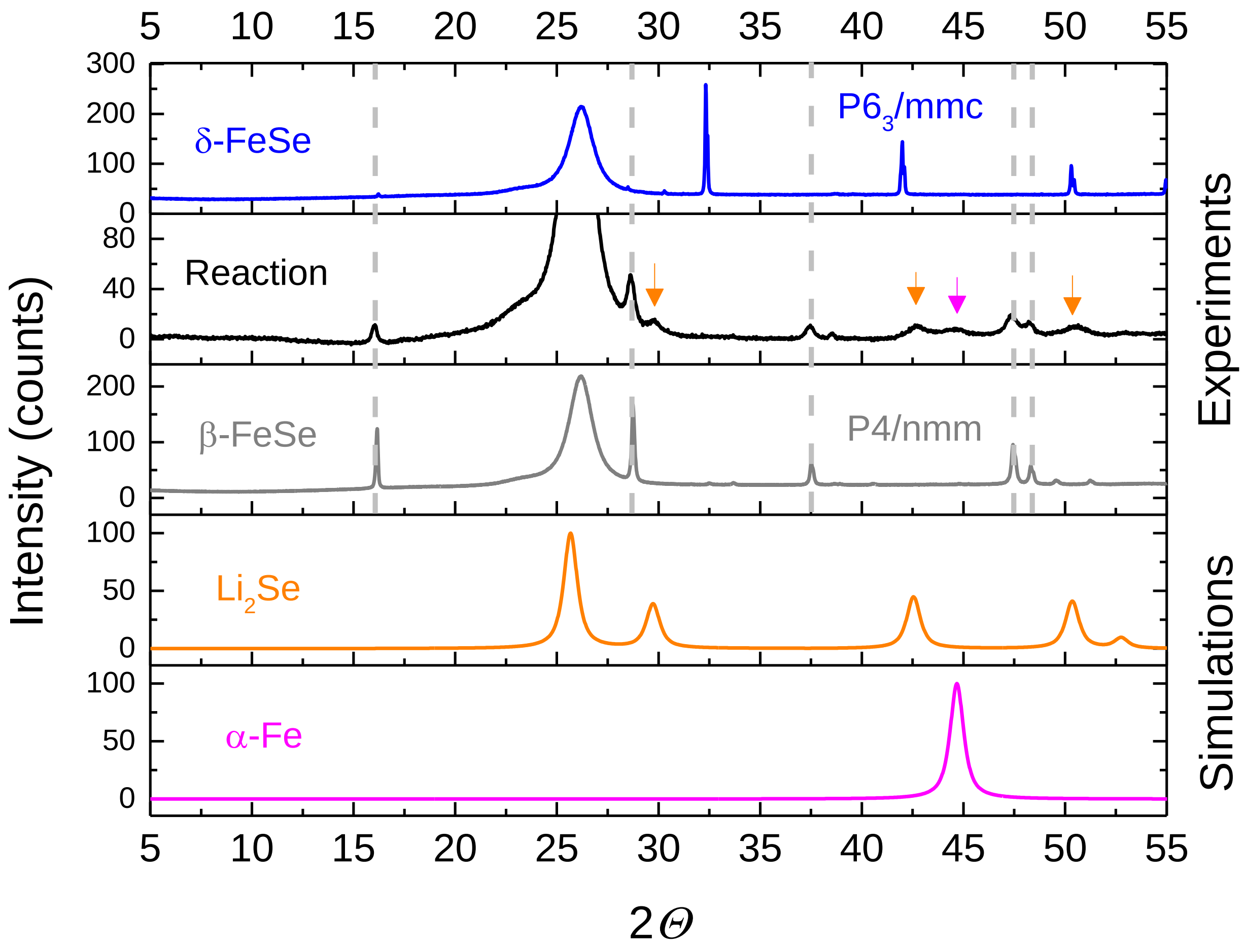

**Figure S8. Ex-situ designed, δ-FeSe phase-interconversion experiments** demonstrate that this high energy Fe-selenide experiences an irreversible decomposition in the "Li-PyH5-FeSe" phase-space. Powder XRDs (Cu, λ = 1.5418 Å) of pure hexagonal *δ*-FeSe (blue line) prepared in advance, was reacted with 0.7 M [Li:$C_5H_5N$] solution at 80°C over a period of 24h (black line). Experimental diffraction pattern of pure tetragonal β-FeSe (grey line) and simulated XRDs for of $Li_2Se$ & metallic *α*-Fe, are plotted for ease of comparison (orange and pink lines, respectively).

## S9. Quantitative atomic PDF analysis

Quantified evaluation of the local structure was successful, primarily for the in-situ experiment (**A**), where a significant intercalated volume fraction (Fig. 4), from the expanded lattice $Li_x(C_5H_5N)_yFe_{2-z}Se_2$ phase, was established. PDFGUI fittings were performed on the basis of a complex mixture of at least three phases, hence lattice constants and thermal parameters of secondary phases, namely, β-FeSe, $\delta$-FeSe, $\alpha$-Fe, and $Li_2Se$ were kept fixed in order to avoid artifacts from cross correlations. Lattice and thermal parameters for β-FeSe, $\delta$-FeSe, and $\alpha$-Fe, were extracted from the multi-phase LeBail and PDFGUI analysis of the starting material (Fig. S9). Tetragonal crystal structure models for the Fe-chalcogenide framework of the starting material (PbFCl-type) and the intercalated phase ($ThCr_2Si_2$-type), both with stoichiometric Fe lattice sites, were utilized in order to further simplify the aforementioned extremely complex fitting process of the in-situ data. The parameter 'detlat1' accounting for correlated atomic motion (cf. nearest-neighbor PDF peak) was fixed equal among all phases, and only 'Qdamp' describing the PDF signal dampening due to particle size was optimized, together with the scale factors of each individual phase, as well as the intercalated phase parameters themselves.

In this process, the PDF data were fitted over a narrow range of interatomic distances, r, probing a length-scale of up to 1 nanometer. The outcomes for selected time-stamps of the experiment (**A**) where the intercalated phase grows to become the majority over time, are compiled in Figure 7. The analysis from fully converged refinements (Fig. 7a; quality of fit factor, $R_w$~10%), allowed us to extract the partial PDFs (Fig. 7b) for the expanded-lattice $Li_x(C_5H_5N)_yFe_{2-z}Se_2$ phase itself, and compare the time-resolved evolution of the $FeSe_4$ local environment (Fig. 8) against that in the parent β-FeSe. With the above knowledge in hand, the validity of the aforementioned multi-phase PDFGUI fitting approach was also implemented in the case of the in-situ experiment (**C**), where however, the expanded lattice $Li_x(C_5H_5N)_yFe_{2-z}Se_2$ phase volume fraction (Fig. S4) did not rich its maximum over the duration of the experiment. The derived parameters, when the Fe-sites retained full stoichiometry, for the experiments (**A**) and (**C**) are summarized in Table S1. Though a reasonably good fit ($R_w$~10%) was obtained from the analysis of the low-r (2-10 Å) G(r) data, a consistent misfit is seen around r ~ 4 Å. In line with previous publications on β-FeSe, [11,12] such a misfit could be accounted for when a local orthorhombic distortion is allowed at temperatures where an average tetragonal symmetry is observed.

Despite the complexity of the multi-parametric G(r) fittings (Fig. 7), assuming several crystallographic phases, a quantifiable estimate of the Fe-site occupancy was possible through the expanded lattice component-phase formed at the different time stamps ($t \geq 6$ h, (**A**)). While there is no significant improvement in the fit quality, when the occupancy of Fe is released in the $ThCr_2Si_2$-type phase (that is a different model than when the Fe-site is kept fully stoichiometric – i.e. at 1 and fixed – Table S1), the analysis suggests a trend where the Fe-site deficiency is raised to ~10% (±1) towards the end of the reaction (25 h).

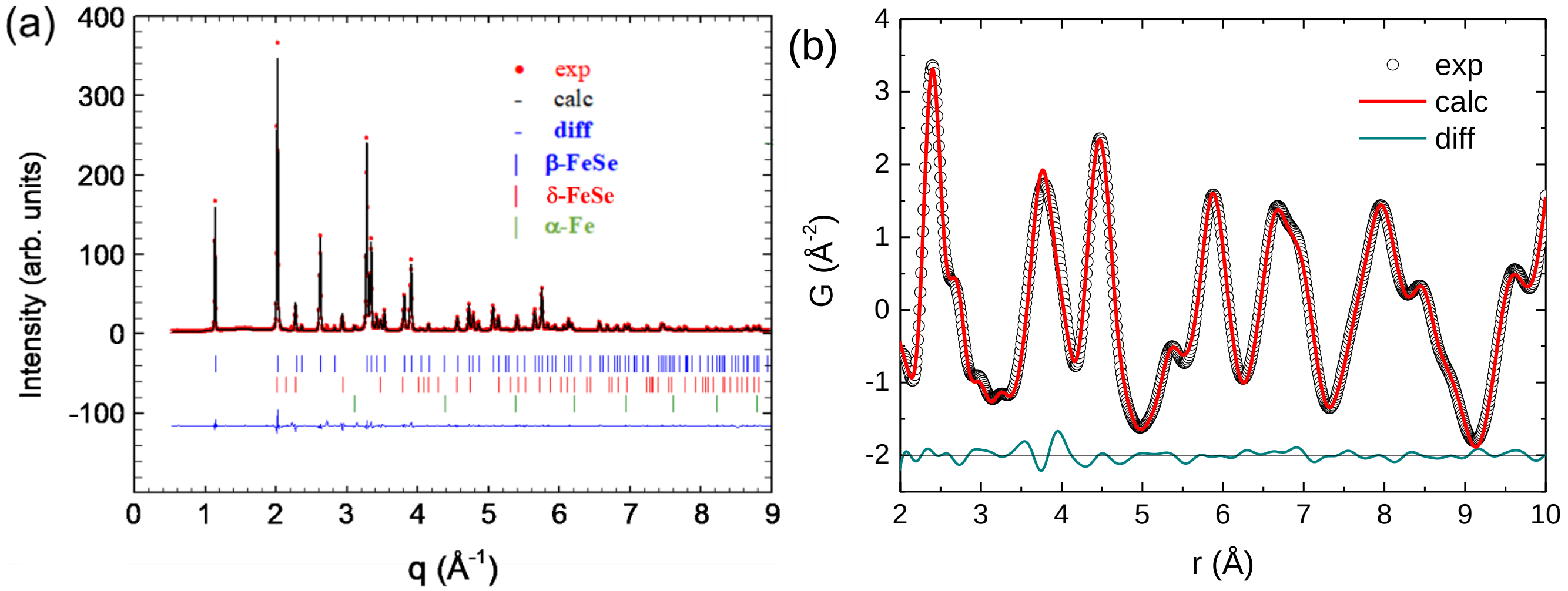

Figure S9. Multi-phase refinements of 300 K synchrotron X-ray scattering data, (a) LeBail fitting, with detector at 'far' side, and (b) PDF fitting, with detector at 'near' side, for a polycrystalline iron selenide starting material, used during the in-situ studies.

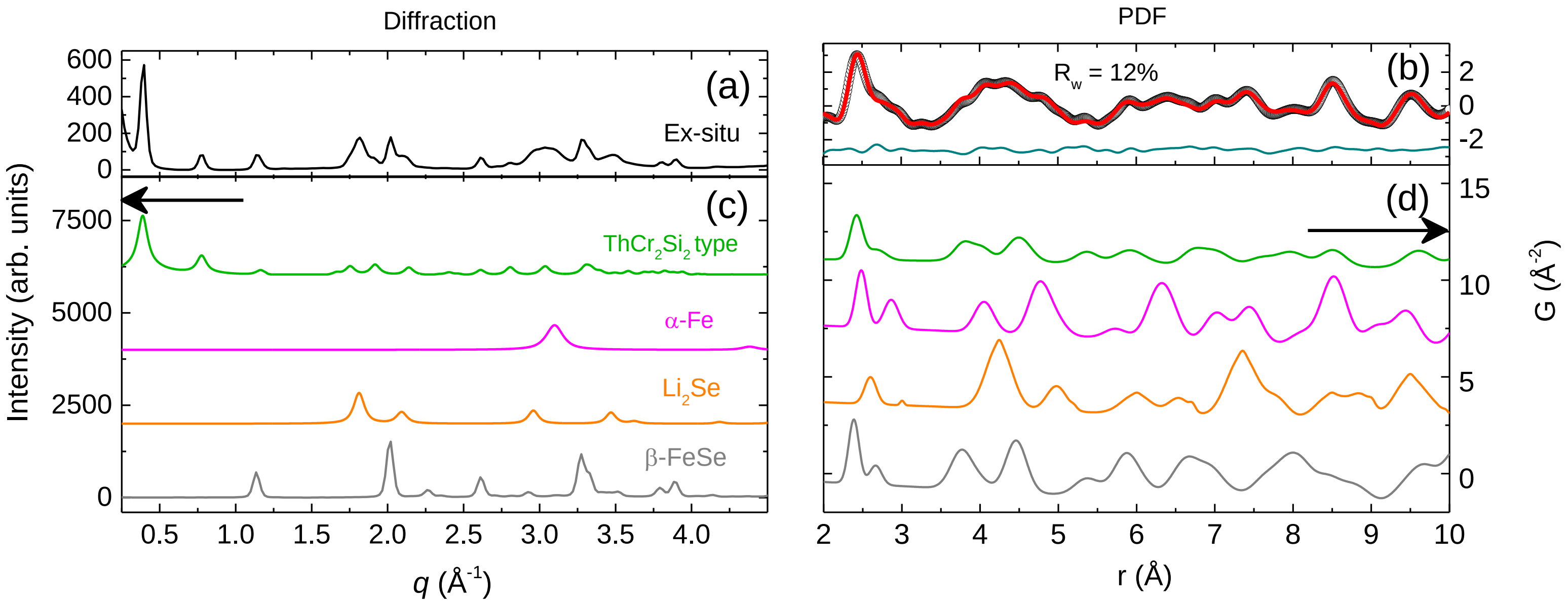

**Figure S10. Synchrotron X-ray total scattering of an ex-situ $Li_x(C_5H_5N)_yFe_{2-z}Se_2$** sample (nominal x~2.0) prepared in advance, with the area detector at near side, for (a) XRD and (b) PDF data collection (λ= 0.1667 Å). (c) Simulations of the powder XRD patterns for the phases involved, namely, the expanded lattice (green), the secondary materials β-FeSe (grey), $Li_2Se$ (orange) *α*-Fe (pink). (d) Partial PDFs derived from fitting (red line) the experimental G(r) (black circles in (b)), with multiple crystallographic phases, including, the intercalated phase (green) described on basis of the $ThCr_2Si_2$ structure type (I4/mmm) and the secondary phases β-FeSe (grey), $Li_2Se$ (orange), *α*-Fe (pink); the difference between experiment and PDF modelling is shown below (dark cyan line).

Table S1. Selected structural parameters reflecting the iron-selenide layer metrics (Å, Å$^3$, °), obtained from PDF analysis of the experimental data. Here, the intercalated phase was analyzed on the basis of the $ThCr_2Si_2$-like crystal type (I4/mmm) and the host β-FeSe, with the PbFCl-like crystal type (P4/nmm), with the occupancy of Fe-sites having retained full stoichiometry in both models. (**A**) and (**C**), reflect outcomes from freshly-made 0.7 M and 72 h aged [Li:PyH5] solutions, respectively.

| Single Layer Metrics | β-FeSe 300K | β-FeSe 350K | $Li_x(C_5H_5N)_yFe_{2-z}Se_2$ *Ex-situ* 300K |
|---|---|---|---|
| Fe−Se (Å) | 2.389(1) | 2.397(0) | 2.420(1) |
| Fe−Fe (Å) | 2.667(0) | 2.673(0) | 2.683(1) |
| a [√2 ×Fe-Fe] (Å) | 3.772(1) | 3.780(0) | 3.795(1) |
| Se−Se (Å) inter | 3.716(4) | 3.711(0) | - |
| Se−Se (Å) | 3.772(1) | 3.780(0) | 3.795(1) |
| Se−Se (Å) | 3.965(2) | 3.980(0) | 4.029(2) |
| Anion height (Å) | 1.467(1) | 1.474(0) | 1.502(1) |
| *α* (°) | 104.2(0) | 104.1(0) | 103.3(1) |
| $V_T$ (Å$^3$), $FeSe_4$ tetrahedron | 6.96(1) | 7.02(0) | 7.21(1) |

| $Li_x(C_5H_5N)_yFe_{2-z}Se_2$ *In-situ* (350K); exp. (**A**) | [~1h] | [~4h] | [~6h] | [~11h] | [~14h] | [~18h] | [~21h] | [~25h] |
|---|---|---|---|---|---|---|---|---|
| Fe−Se (Å) | - | 2.404(1) | 2.409(5) | 2.418(5) | 2.406(0) | 2.409(1) | 2.408(1) | 2.414(1) |
| Fe−Fe (Å) | - | 2.629(1) | 2.630(10) | 2.656(3) | 2.665(2) | 2.680(2) | 2.686(3) | 2.706(2) |
| a [√2 ×Fe-Fe] (Å) | - | 3.718(1) | 3.720(10) | 3.756(4) | 3.769(3) | 3.791(3) | 3.799(4) | 3.827(3) |
| Se−Se (Å) | - | 3.718(2) | 3.720(20) | 3.756(4) | 3.769(2) | 3.791(3) | 3.799(4) | 3.827(3) |
| Se−Se (Å) | - | 4.025(1) | 4.037(5) | 4.040(10) | 4.006(0) | 4.004(0) | 3.998(1) | 3.997(0) |
| Anion height (Å) | - | 1.524(1) | 1.531(2) | 1.524(9) | 1.495(1) | 1.487(1) | 1.481(1) | 1.470(1) |
| *α* (°) | - | 101.3(0) | 101.1(3) | 101.9(4) | 103.1(1) | 103.8(1) | 104.1(1) | 104.9(1) |
| $V_T$ (Å$^3$), $FeSe_4$ tetrahedron | - | 7.02(1) | 7.07(5) | 7.17(3) | 7.08(1) | 7.12(1) | 7.12(1) | 7.18(1) |

| $Li_x(C_5H_5N)_yFe_{2-z}Se_2$ *In-situ* (350K); exp. (**C**) | [~9h] | [~12h] | [~15h] | [~18h] | [~21h] |
|---|---|---|---|---|---|
| Fe−Se (Å) | 2.425(1) | 2.420(1) | 2.416(1) | 2.412(1) | 2.415(0) |
| Fe−Fe (Å) | 2.641(1) | 2.667(7) | 2.690(3) | 2.678(1) | 2.691(2) |
| a [√2 ×Fe-Fe] (Å) | 3.735(1) | 3.770(10) | 3.805(4) | 3.787(1) | 3.806(3) |
| Se−Se (Å) | 3.735(2) | 3.770(10) | 3.805(4) | 3.787(1) | 3.806(3) |
| Se−Se (Å) | 4.069(1) | 4.039(4) | 4.014(0) | 4.011(0) | 4.010(3) |
| Anion height (Å) | 1.547(1) | 1.516(6) | 1.490(2) | 1.493(0) | 1.486(1) |
| *α* (°) | 100.7(4) | 102.4(4) | 103.9(1) | 103.5(0) | 104.0(1) |
| $V_T$ (Å$^3$), $FeSe_4$ tetrahedron | 7.20(1) | 7.19(1) | 7.19(1) | 7.14(0) | 7.18(1) |